\documentclass[twocolumn]{aastex631}
\usepackage{natbib}
\usepackage{amsmath,amssymb, hyperref}
\usepackage[shortlabels]{enumitem}
\usepackage{booktabs}
\defcitealias{Sternberg1989a}{S89}

\shorttitle{Constraining H$_2$ formation and dissociation rates with observations}
\shortauthors{Bialy et al. }

\begin{document}

\title{The Molecular Cloud Lifecycle I: 
\\
Constraining H$_2$ formation and dissociation rates with observations}

\correspondingauthor{Shmuel Bialy}
\author[0000-0002-0404-003X]{Shmuel Bialy}
\affiliation{Physics Department, Technion - Israel Institute of Technology, Haifa 32000, Israel}

\author{Blakesley Burkhart}
\affiliation{Department of Physics and Astronomy, Rutgers University,  136 Frelinghuysen Rd, Piscataway, NJ 08854, USA}
\affiliation{Center for Computational Astrophysics, Flatiron Institute, 162 Fifth Avenue, New York, NY 10010, USA}

\author{Daniel Seifried}
\affiliation{I. Physics Institute, University of Cologne, Z\"ulpicher Str. 77, 50937 Cologne, Germany}
\affiliation{Center for Data and Simulation Science, www.cds.uni-koeln.de, Cologne, Germany}

\author{Amiel Sternberg}
\affiliation{Tel Aviv University, P.O. Box 39040, Tel Aviv 6997801, Israel}
\affiliation{Center for Computational Astrophysics, Flatiron Institute, 162 Fifth Avenue, New York, NY 10010, USA}

\author{Benjamin Godard}
\affiliation{Observatoire de Paris, Universite PSL, Sorbonne Universite, LERMA, 75014 Paris, France}
\affiliation{Laboratoire de Physique de l’Ecole Normale Superieure, ENS, Universite PSL, CNRS, Sorbonne Universite, Universite de Paris, F-75005 Paris, France}

\author{Mark R. Krumholz}
\affiliation{Research School of Astronomy and Astrophysics, Australian National University, Canberra ACT 2600 Australia}

\author{Stefanie Walch}
\affiliation{I. Physics Institute, University of Cologne, Z\"ulpicher Str. 77, 50937 Cologne, Germany}
\affiliation{Center for Data and Simulation Science, www.cds.uni-koeln.de, Cologne, Germany}

\author[0000-0002-3131-7372]{Erika Hamden}
\affiliation{Steward Observatory, University of Arizona, Tucson, AZ 85719, USA}

\author{Thomas J. Haworth}
\affiliation{Astronomy Unit, School of Physics and Astronomy, Queen Mary University of London, London E1 4NS, UK}

\author{Neal J. Turner}
\affiliation{Jet Propulsion Laboratory, California Institute of Technology, Pasadena, CA 91109, USA}

\author{Min-Young Lee}
\affiliation{Korea Astronomy and Space Science Institute, 776 Daedeok-daero, Daejeon 34055, Republic of Korea}

\author{Shuo Kong}
\affiliation{Steward Observatory, University of Arizona, Tucson, AZ 85719, USA}

\email{sbialy@technion.ac.il}

\begin{abstract}

Molecular clouds (MCs) are the birthplaces of new stars in galaxies. A key component of MCs are photodissociation regions (PDRs), where far-ultraviolet radiation plays a crucial role in determining the gas's physical and chemical state. Traditional PDR models assume chemical steady state (CSS), where the rates of H$_2$ formation and photodissociation are balanced. However, real MCs are dynamic and can be out of CSS.
In this study, we demonstrate that combining H$_2$ emission lines observed in the far-ultraviolet or infrared with column density observations can be used to derive the rates of H$_2$ formation and photodissociation. We derive analytical formulae that relate these rates to observable quantities, which we validate using synthetic H$_2$ line emission maps derived from the SILCC-Zoom hydrodynamical simulation. 
 Our method estimates integrated H$_2$ formation and dissociation rates with an accuracy $\approx30$\% (on top of uncertainties in observed H$_2$ emission maps and column densities). Our simulations, valid for column densities $N \leq 2 \times 10^{22}$ cm$^{-2}$,
 cover a wide dynamic range in H$_2$ formation and photodissociation rates, showing significant deviations from CSS, with 74\% of the MC's mass deviating from CSS by a factor greater than 2. 
Our analytical formulae can effectively distinguish between regions in and out of CSS. When applied to actual H$_2$ line observations, our method can assess the chemical state of MCs, providing insights into their evolutionary stages and lifetimes. A NASA Small Explorer mission concept, \textit{Eos}, will be proposed in 2025 and is specifically designed to conduct the types of observations outlined in this study.

\end{abstract}

\section{Introduction}
\label{sec: intro}
Molecular hydrogen (H$_2$), the most abundant molecule in the universe, plays a crucial role in the lifecycle of baryons throughout cosmic history \citep{Galli1998a, McKee2007, tacconiEvolutionStarFormingInterstellar2020}. It acts as a vital cooling agent in the early universe \citep{Haiman1996, Bromm2001, Omukai2000, Barkana2001, Bialy2019}, triggers rich chemistry in the interstellar medium (ISM) \citep{Herbst1973, Tielens2013, VanDishoeck2013a, Bialy2015a}, and correlates with the star formation rate in present-day galaxies \citep{Bigiel2008, Leroy2008, Schruba2011}.

H$_2$ is predominantly found in molecular clouds (MCs) and is excited by far-UV (FUV) radiation within the Lyman-Werner (LW) band ($h \nu = 11.2-13.6$ eV). 
LW photons excite the electronic $B^1\Sigma^+_u$ and $C^1\Pi_u$ states of H$_2$. These excited states can then radiatively decay through two pathways: (a) to the rovibrational continuum, leading to H$_2$ dissociation and the emission of FUV continuum radiation; or (b) to a bound (rovibrationally excited) level in the ground electronic state ($X^1\Sigma^+_g$), accompanied by FUV line emission
(\citealt{Field1966, stecher1967, Dalgarno1970, Sternberg1989a}, hereafter \citetalias{Sternberg1989a}).
The rovibrationally excited H$_2$ molecules continue to cascade down the rovibrational ladder, emitting infrared (IR) lines \citep{Black1987, Sternberg1988, Luhman1994, Goldshmidt1995a, Draine1996, Neufeld1996, Le2017, Kaplan2021}.

In this study, we focus on photodissociation regions (PDRs), where radiative processes dominate molecular excitation and dissociation \citep{Tielens1985, Hollenbach1999, LePetit2006, rolligAstrophysicsPhotonDominated2007, Bisbas2021, rollig2022, Pound2023}. This includes gas in the vicinity of massive stars (e.g., the Orion nebula) and MCs embedded in the ambient interstellar radiation field, as long as the dust visual extinction is not too large. In contrast, in cloud cores, LW radiation is strongly attenuated due to H$_2$ self-shielding and dust absorption. In these well-shielded regions, H$_2$ excitation and dissociation are driven by deep-penetrating cosmic rays or X-rays \citep{Maloney1996, Dalgarno2006, Wolfire2022, Bialy2020}. However, even for MCs exposed only to the ambient interstellar radiation field, photoprocesses dominate H$_2$ dissociation, excitation, and line emission up to gas column densities of $N \approx 10^{22}$ cm$^{-2}$ \citep[][see also \S \ref{sub: discuasion - CRs} of this paper, and Appendix \ref{app: CRs}]{Sternberg2024, Padovani2024}.

Traditional PDR models assume that the total formation and destruction rates of H$_2$ (and other molecules) balance each other, i.e., that the system is in chemical steady state (CSS). This allows efficient characterization of MCs since the abundance and emission spectrum of H$_2$ are time-independent and depend only on physical conditions such as the FUV radiation field intensity, gas density, and gas metallicity. For example, the CSS assumption allows the derivation of useful analytic formulae describing: (a) the atomic-to-molecular transition point \citep{Bialy2016a}, (b) the total HI column density \citep{Krumholz2008, Krumholz2009, Sternberg2014, Bialy2017}, and (c) the total intensity of H$_2$ line emission (\citetalias{Sternberg1989a}). This analytic framework has been utilized in the analysis of observations in various Galactic and extragalactic PDRs \citep[e.g.][]{Bialy2015c, Schruba2018, Ranjan2018, Noterdaeme2019, syed2022}.

However, in practice, the assumption of CSS may be problematic. MCs are dynamic entities. They form in regions of converging flows (e.g., in gas that is infalling onto galactic spiral arms, collisions of expanding supernova shells, etc.), where gas may be compressed to high densities \citep{koyama2000, Hartmann2002, Ntormousi2011, dawson2013, bialy2021}. Once gravitational collapse is initiated, newly formed stars begin to disperse the gas through various stellar feedback processes: ionizing radiation, stellar winds, jets, and supernova explosions \citep{McKee1977, Faucher-Giguere2013b, hopkins2020, Orr2022, ostriker2022, Chevance2023}. These dynamical processes can occur on short timescales compared to the time required for the gas to achieve CSS, and thus MCs may be out of CSS \citep{Glover2007a, Krumholz2012a, Richings2014b, Hu2016, Valdivia2016, Seifried2017, Seifried2021}.

This raises an important question:
{\bf Can we determine from observations whether a given MC is in or out CSS?}

In this paper, we illustrate how combining the total intensity in H$_2$ line emission with the total gas and HI column densities along the line of sight (LOS), allows us to obtain reliable estimates for the column-integrated rates of H$_2$ photodissociation and formation. This enables us to assess whether a CSS is maintained.

The question of whether a MC is in or out of CSS has important implications for MCs' lifetimes and evolution.
If the gas in a MC is found to be far from CSS it implies that the MC has either been recently replenished with ``fresh" gas or has lost gas via evaporation, over a timescale that is short compared to the chemical time (i.e., Eq.~\ref{eq: t chem} below; see also \citealt{Jeffreson2024}).
In a second paper in this series \citep{Burkhart2024}, we explore in more detail the time evolution of MCs,
the evolution of the H$_2$ formation and photodissociation rates in MCs, and their relationship 
to the star formation rate.

The structure of this paper is as follows:
in \S \ref{sec: theory}, we provide the fundamental theoretical framework. We derive the key analytical equations, namely Eqs.~(\ref{eq: D-Itot}, \ref{eq: JF observation 2}), which elucidate how an observer can employ H$_2$ line emission intensities and column density maps to calculate integrated H$_2$ formation and dissociation rates along the line of sight.
In \S \ref{sec: Numerical Model} we present the magnetohydrodynamical simulations and the numerical procedure for producing H$_2$ line intensity maps.
We use these maps to test our analytic theory.
In \S \ref{sec: results} we present our results, relating the state of the gas in the simulation and the H$_2$ formation and dissociation rates in various cloud regions to the observables.
We follow up with a discussion and conclusions in \S \ref{sec:dis} and \S \ref{sec:con}.

\section{Theoretical Model}
\label{sec: theory}

\subsection{H$_2$ formation and dissociation}
\label{sub: H2 formation dissociation}
For typical ISM conditions, H$_2$ formation is dominated by dust catalysis \citep{Wakelam2017}. 
The destruction of H$_2$ is dominated by photodissociation.
The net change in number of H$_2$ molecules per unit time and volume is
\begin{equation}
\label{eq: dnH2_dt}
    \frac{\mathrm{d}n({\rm H_2})}{\mathrm{d}t}=j_F-j_D
\end{equation}
where
\begin{align}    
\label{eq: jf jd}
    j_F \equiv n({\rm H}) n R \; , \; \;
    j_D \equiv n({\rm H_2}) D \ ,
\end{align}
are the volumetric H$_2$ formation and dissociation rates (cm$^{-3}$ s$^{-1}$).
Here, $R$ (cm$^3$ s$^{-1}$) is the H$_2$ formation rate coefficient, $D$ (s$^{-1}$) is the local photodissociation rate, which may be significantly attenuated due to H$_2$ self-shielding and dust absorption (see below), $n({\rm H})$ and $n({\rm H_2})$, are the H and H$_2$ number densities, respectively, and $n$ is the total hydrogen nucleon density (cm$^{-3}$). 
In Eq.~(\ref{eq: jf jd}) and throughout our analytic model we consider only H$_2$ photodissociation and neglect additional H$_2$ destruction via cosmic-rays. 
This assumption is justified in \S \ref{sub: discuasion - CRs} and in Appendix \ref{app: CRs}.

In this work, we use the ``SImulating the Life-Cycle of molecular Clouds" (SILCC)-Zoom \citep{Seifried2017} simulation suite to produce synthetic maps of H$_2$ line emission (as described below).
In line with the SILCC simulation suite, we adopt an H$_2$ formation rate coefficient
\begin{equation}
\label{eq: R}
    R = 3 \times 10^{-17} T_2^{1/2} S f_a Z_d' \ {\rm cm^3 \ s^{-1}} \ ,
\end{equation}
 where $S = [1 + 0.4 (T_2+T_{d,2})^{0.5} + 0.2 T_2 + 0.08 T_2^2]^{-1}$
is the sticking coefficient, $f_a = [1 + 10^4 \mathrm{e}^{-600 {\rm K} / T_d}]^{-1}$
is the fraction of H atoms that enter the potential wells on the dust grain before evaporating and thus ultimately combine to form H$_2$ molecules \citep{Hollenbach1979}, $Z_d'$ is the dust-to-gas ratio relative to the Solar neighborhood ISM, $T$ and $T_d$ are the gas and dust temperatures, and $T_{2} \equiv T/(10^2 {\rm K})$.
The H$_2$ photodissociation rate is given by
\begin{align}
\label{eq: D}
    D = \chi D_0 f_{\rm H_2, shield} f_{\rm dust}  \ ,
\end{align}
where $\chi=F_{\rm FUV}/F_{0}$ is the flux of the incident FUV radiation field on the cloud relative to the typical Solar neighborhood value, $F_0=2.7 \times 10^{-3}$ erg cm$^{-2}$ s$^{-1}$ 
\citep[][]{Draine1978, Bialy2020}, and $D_0=5.8 \times 10^{-11}$ s$^{-1}$ is the free-space photodissociation rate in the absence of shielding \citep{Sternberg2014}. 
The functions $f_{\rm H_2,  shield}$ and $f_{\rm dust}$ account for LW attenuation by H$_2$ lines (self-shielding) and by dust absorption (see \S \ref{sub:num} and the discussion after Eqs.~\ref{app eq: Itot 1}-\ref{app eq: Itot 2} for more details).

In CSS $j_F=j_D$, and at any cloud position, the H$_2$-to-H ratio is then given by $n({\rm H_2})/n({\rm H}) = Rn/D$.
If $j_F \neq j_D$ the system is out of CSS: for $j_F > j_D$, there is net H$_2$ formation and the H$_2$ mass grows with time, whereas if $j_F < j_D$ the H$_2$ mass decreases with time.

For a given value of $D$, $R$ and $n$, the timescale to reach CSS is 
\begin{equation}
\label{eq: t chem}
    t_{\rm chem} = \frac{1}{2nR+D} \ ,
\end{equation}
This follows from Eq.~(\ref{eq: dnH2_dt}-\ref{eq: jf jd}) \citep[see ][for a discussion of this and other relevant timescales. See also \citealt{Goldshmidt1995a, Goldsmith2007}]{Bialy2017}.
Near cloud boundaries, where there is no LW attenuation, $D=\chi D_0 \gg 2 R n$. Under these conditions, the chemical time is very short: $t_{\rm chem}=1/(\chi D_0)=550/\chi$ years.
In deep cloud interiors, where radiation is significantly attenuated by dust and line absorption, $D \ll 2Rn$.
 In this regime, the chemical time equals the H$_2$ formation time:
\begin{equation}
    \label{eq: t H2 form}
    t_{\rm chem} \rightarrow t_{\rm H_2, form} \equiv \frac{1}{2Rn} 
    \approx
    9 \left( \frac{1}{Z_d' n_2}\right) \ {\rm Myr} \ ,
\end{equation}
where for the numerical evaluation we 
used Eq.~(\ref{eq: R}) with typical CNM conditions, $T=100$ K, $T_d \ll T$, and defined $n_2 \equiv n/(100 \ {\rm cm^{-3}})$.
Thus, cloud envelopes, characterized by short chemical timescales, tend to be in CSS, while cloud interiors, which exhibit long chemical timescales, are prone to deviate from CSS.

Hereafter we adopt typical solar-neighborhood values for the FUV radiation field intensity and the dust-to-gas ratio, 
$Z_d' = 1$, $\chi = 1$.

\subsection{Estimating H$_2$ formation-dissociation with observations}
\label{sub: H2 formation dissociation - obs}

In this subsection, we discuss how we can use emission line observations to derive the H$_2$ formation and dissociation rates.
As observations are probing integral quantities (integrated along the LOS) rather than volumetric quantities, we define the column-integrated H$_2$ formation and dissociation mass rates
\begin{align}
\label{eq: JF JD}
    \dot{\Sigma}_{F}^{\rm (true)} &\equiv \bar{m} \int j_F \mathrm{d}s = \bar{m} \int n({\rm H}) n R \mathrm{d}s \\ \nonumber
    \dot{\Sigma}_{D}^{\rm (true)} &\equiv \bar{m} \int j_D \mathrm{d}s = \bar{m} \int n({\rm H_2}) D \mathrm{d}s \ ,
\end{align}
where $s$ is the coordinate along the LOS.
The quantities $\dot{\Sigma}_{F}^{\rm (true)}$ and $\dot{\Sigma}_{D}^{\rm (true)}$ express the gas mass that is converted from atomic to molecular form and vise versa, per unit area and time ($M_{\odot}$ pc$^{-2}$  Myr$^{-1}$).
%Here, $\dot{\Sigma}_{F}^{\rm (true)}$ and $\dot{\Sigma}_{D}^{\rm (true)}$ are integrated rates.
We adopt a mean particle mass $\bar{m}=2 m_{\rm H} \times 1.4 = 4.7 \times 10^{-24}$ g, corresponding to the mass of an H$_2$ molecule, with the additional helium contribution assuming cosmic He abundance.   
We use the superscript ``(true)'' to stress that these are the true rates, as calculated by integrating the volumetric rates in our simulation.
This is as opposed to the observationally-estimated rates (defined below), that are derived from  observable quantities such as H$_2$ line intensities and column densities.

\subsubsection{The H$_2$  photodissociation rate}
\label{sub: H2 dissociation - obs}

As discussed in \S \ref{sec: intro}, H$_2$ photodissociation occurs via a two-step process in which first the electronic states of H$_2$ are photo-excited. 
The radiative decay to the rovibrational continuum of the ground electronic state leads to H$_2$ dissociation.
The probability of dissociation per excitation is given by
\begin{equation}
    \label{eq: p_diss}
    p_{\rm diss} \equiv {D/P} \approx 0.15 \ ,
\end{equation}
where $P$ is the total H$_2$ photo-excitation rate (of all H$_2$ electronic states) \citep{Abgrall1992, Draine1996}.
In the remaining 85\% of the cases, the H$_2$ decays to rovibrational bound states, producing FUV and subsequently IR line emission.
In addition to line emission, H$_2$ electronic excitation also results in the emission of continuum FUV radiation \citep{Dalgarno1970} which can also be observed and used to constrain the H$_2$ excitation and dissociation rates. 
In this paper, we focus on line emission, with our analytic and numerical analysis in \S\S  \S \ref{sec: theory}-\ref{sec: results} concentrating on FUV lines. We then generalize to IR lines in \S \ref{sec:dis}.

Since the process of H$_2$ photo-excitation results in both line emission and H$_2$ dissociation, the H$_2$ dissociation rate is proportional to the total intensity of the H$_2$ emission lines.
As we show in the Appendix (Eq.~\ref{app eq: D-Itot}), this relation is given by
\begin{align}
\label{eq: D-Itot}
    \dot{\Sigma}_{D}^{\rm (obs)} &= \frac{4 \pi p_{\rm diss} \bar{m} }{1-p_{\rm diss}} \ \mathcal{I}_{\rm tot} \left(\frac{\tau_{\rm tot}}{1-\mathrm{e}^{-\tau_{\rm tot}}} \right)  \\ \nonumber 
    &= 
    0.30 \  \mathcal{I}_5 \left( \frac{N_{21}}{1-\mathrm{e}^{-1.9 N_{21}}} \right)  \ {\rm M_{\odot}  \ pc^{-2} \ Myr^{-1}} \ ,
\end{align}
where $\mathcal{I}_{\rm tot}$ is the total photon intensity summed over all the FUV emission lines (photons cm$^{-2}$ s$^{-1}$ str$^{-1}$), and 
$\tau_{\rm tot}$ is the dust opacity in the LW band.
 In Eq.~\ref{eq: D-Itot} we defined $\mathcal{I}_{\rm 5} \equiv \mathcal{I}_{\rm tot}/({10^{5} \ {\rm photons \ cm^{-2} \ s^{-1} \ sr^{-1}}})$ and used
 $\tau_{\rm tot}=\sigma N = 1.9 N_{21}$ where $\sigma =1.9 \times 10^{-21}$ cm$^2$ \citep{Sternberg2014}
 is the dust absorption cross section per hydrogen nucleus, $N$ is the column density of hydrogen nuclei along the LOS and $N_{21} \equiv N/(10^{21} \ {\rm cm^{-2}})$.

In Eq.~(\ref{eq: D-Itot}), we use the superscript (obs) to indicate that this expression approximates the true photodissociation rates, relying on integrated observable quantities rather than the detailed 3D density structure, and radiation geometry information (see Appendix \ref{app: Sigma-Itot}).
We note that in practice, observers typically measure only a subset of H$_2$ lines, rather than the total line emission $\mathcal{I}_{\rm tot}$. However, theoretical methods exist to estimate $\mathcal{I}_{\rm tot}$ from a subset of observed lines using robust line ratios that are relatively insensitive to physical conditions \citep{Black1987, Sternberg1989a}. The detailed methodology for this conversion will be explored in future work.

The factor in parenthesis expresses the absorption of the emitted H$_2$ lines by intervening dust.
This factor connects smoothly the optically thin and thick regimes.
In the optically thin limit ($\tau_{\rm tot} \ll 1$) this factor approaches unity, and $\dot{\Sigma}_{D}^{\rm (obs)} \propto \mathcal{I}_{\rm tot}$. In this limit, the H$_2$ emission lines directly trace the integrated H$_2$ photodissociation rate. 
On the other hand, in the optically thick limit ($\tau_{\rm tot} > 1$), the factor $\tau_{\rm tot}/(1-\mathrm{e}^{-\tau_{\rm tot}}) \rightarrow \tau_{\rm tot}$, and the ratio $\dot{\Sigma}_{D}^{\rm (obs)}/\mathcal{I}_{\rm tot}$ grows linearly with $\tau_{\rm tot}$. 
In this limit, $\mathcal{I}_{\rm tot}$ traces only the outer part of the cloud, at an optical depth of $\approx 1$. While observed lines originate mainly from outer cloud envelopes, H$_2$ photodissociation can still occur deeper within clouds as FUV radiation penetrates through lower opacity regions in the patchy structure, not necessarily along the LOS.
% Overall, the $\tau_{\rm tot}/(1-\mathrm{e}^{-\tau_{\rm tot}})$ factor accounts for this reduction due to the finite optical thickness (see Appendix).

For typical molecular clouds in our Galactic neighbourhood, $\tau_{\rm tot} \approx 1$ ($N_{21} \approx 0.5$), $\mathcal{I}_5 \approx 0.3-0.6$\footnote{This follows from scaling \citetalias{Sternberg1989a}'s results to $\chi=1$. \citetalias{Sternberg1989a} obtained a total H$_2$ line intensity of $1.1 \times 10^{-4}$ erg cm$^{-2}$ s$^{-1}$ sr$^{-1}$ for his fiducial $\chi=100$, $n=10^3$ cm$^{-3}$ model. As discussed in \citetalias{Sternberg1989a}, these parameters correspond to the ``weak-field`` limit in which $\mathcal{I}_{\rm tot} \propto \chi$.
Scaling \citetalias{Sternberg1989a}'s result to $\chi=1$ and dividing by a mean photon energy $\langle h \nu \rangle = 11.2$ eV we obtain $\mathcal{I}_5 = 0.6$.}, and the integrated H$_2$ photodissociation rate is 
$\approx 0.1$ M$_{\sun}$ pc$^{-2}$ Myr$^{-1}$.

\subsubsection{The H$_2$ formation rate}
Since H$_2$ formation involves H atoms that interact on dust grains, the integrated formation rate may be derived from observations of the H {\small I} column density (via the 21 cm emission line).
To see this, we first define the effective-mean $R n$ factor,
\begin{equation}
\label{eq: nR eff}
    \langle Rn \rangle_{\rm eff} \equiv \frac{\int (Rn) n({\rm H}) \mathrm{d}s}{\int n({\rm H}) \mathrm{d}s} \ .
\end{equation}
Using this definition, we can rewrite Eq.~(\ref{eq: JF JD}) as
% $\dot{\Sigma}_{F}^{\rm (obs)} = \bar{m} \langle Rn \rangle_{\rm eff} N({\rm H})$,
\begin{equation}
% \label{eq: JF observation}
    \dot{\Sigma}_{F}^{\rm (true)} = \bar{m} \langle Rn \rangle_{\rm eff} N({\rm H}) \ ,
\end{equation}
where $N({\rm H})$ is the H column density.

First, let us gain intuition by considering a simple case of a uniform density and temperature slab.
In this case, Eq.~(\ref{eq: nR eff}) simplifies to   $\langle Rn \rangle_{\rm eff} = Rn$.
For typical cold neutral medium (CNM) conditions $n\approx 30$ cm$^{-3}$, $T=100$ K \citep{Wolfire2003, Bialy2019}. With Eq.~(\ref{eq: R})
we get 
$Rn = 5.4 \times 10^{-16}$ s$^{-1}$. 
For a CNM column density of $5 \times 10^{20}$ cm$^{-2}$ we get an integrated H$_2$ formation rate of
$\approx 0.2~{\rm M_{\odot} \ pc^{-2} \ Myr^{-1}}$.
% $\dot{\Sigma}_{F}= 10^5$ cm$^{-2}$ s$^{-1}$.

In practice, the gas density and temperature vary inside the cloud, and the value of $\langle Rn \rangle_{\rm eff}$ differs from one LOS to another. 
3D dust maps offer insights into density structure (e.g., \citealt{Leike2020a, Zucker2021}), but they lack the resolution to capture the critical HI-to-H$_2$ transition length. This transition, occurring over scales $\lesssim 1$ pc \citep{Bialy2017}, is crucial for studying the H-H$2$ balance and clouds' chemical state. Due to this resolution limitation, we must rely on readily observable LOS-integrated quantities to estimate $\langle Rn \rangle{\rm eff}$.
From a theoretical point of view, 
$\langle R n \rangle_{\rm eff}$ is expected to correlate with the integrated gas column density $N$.
This is because particles that are situated in large reservoirs of mass will typically have large integrated column density along the LOS (i.e., large $N$), while on the other hand these particles are situated in deeper gravitational potential wells leading to gas compression (i.e., higher $n$).

This correlation has been observed in various independent hydro simulations of the interstellar medium \citep[i.e., the $A_V-n$ relation;][]{Bisbas2019, huMetallicityDependenceCO2021, Bisbas2021, gaches2022}.
Using our SILCC-Zoom simulations, we  find that $\langle Rn \rangle_{\rm eff}$ is well described by the powerlaw relation
$\langle Rn \rangle_{\rm eff} = k_0 N_{21}^{\alpha}$
with $k_0 = 2.0 \times 10^{-16}$ s$^{-1}$ and $\alpha=1.3$.
% (see \S\ref{sub: synthtic obs} for details).
With this power law relation we get
\begin{align}
\label{eq: JF observation 2}
    \dot{\Sigma}_{F}^{\rm (obs)} &= \bar{m} k_0 N_{21}^{\alpha} N(\rm H)\\ \nonumber 
    &= 0.14 \ f_{\rm H} N_{21}^{1+\alpha} \ {\rm M_{\sun} \ pc^{-2} \ Myr^{-1}} \ ,
\end{align}
where $\alpha=1.3$, and where we defined
$f_{\rm H} \equiv N({\rm H})/N$. 

Similarly to the case of $\dot{\Sigma}_{D}^{\rm (obs)}$, here, too, we use the superscript (obs) to emphasize that $\dot{\Sigma}_{F}^{\rm (obs)}$ is derived based on observed quantities only, and is an approximation to the true rate.
In \S \ref{sec: results} we test the accuracy of these approximations using our hydro simulations. 
% Observational tests are possible in specific scenarios. When H~\textsc{i} is observed in absorption against a background radio continuum source or in self-absorption, we can estimate the mass fraction and volume density of the cold H~\textsc{i} phase \citep{McClure-Griffiths23a}. This cold phase likely dominates H$_2$ formation. However, conducting such observational tests falls outside the scope of our current study.

\subsubsection{The formation to photodissociation rate ratio}
Combining Eqs.~(\ref{eq: D-Itot}) and (\ref{eq: JF observation 2}),  we obtain  the formation to dissociation rate ratio
\begin{align}
\label{eq: JF_to_JD}
    \left( \frac{\dot{\Sigma}_{F}}{\dot{\Sigma}_{D}} \right)^{\rm (obs)} 
    &= \frac{k_0 N_{21}^{\alpha} N({\rm H})}{4 \pi \mathcal{I}_{\rm tot}  (\frac{\tau_{\rm tot}}{1-\mathrm{e}^{-\tau_{\rm tot}}})} \frac{1-p_{\rm diss}}{p_{\rm diss}}\ , \\ \nonumber
    &= 0.47 \ N_{21}^{\alpha} f_{\rm H} \left( 1-\mathrm{e}^{-1.9 N_{21}} \right) \mathcal{I}_5^{-1} \ .
\end{align}
% For instance, for $N_{21}=1$, $f_{\rm H}=0.5$, $\mathcal{I}_5=0.3$ we get a ratio of $0.8$, i.e., a situation where the formation and dissociation rates approximately balance each other.

If CSS holds, $\dot{\Sigma}_{F}/\dot{\Sigma}_{D}=1$.
If $\dot{\Sigma}_{F}/\dot{\Sigma}_{D} > 1$, then the gas along the LOS is not in CSS, and the H$_2$ column density increases 
with time at the expense of HI.
If $\dot{\Sigma}_{F}/\dot{\Sigma}_{D} < 1$, the H$_2$ column density decreases with time, and HI increases.
Thus, given an observation of the gas HI and total column density and an H$_2$ emission spectrum, we can constrain the chemical state of the gas and whether it is in CSS, or not.

\begin{figure*}
    \centering
    \includegraphics[width=1\textwidth]{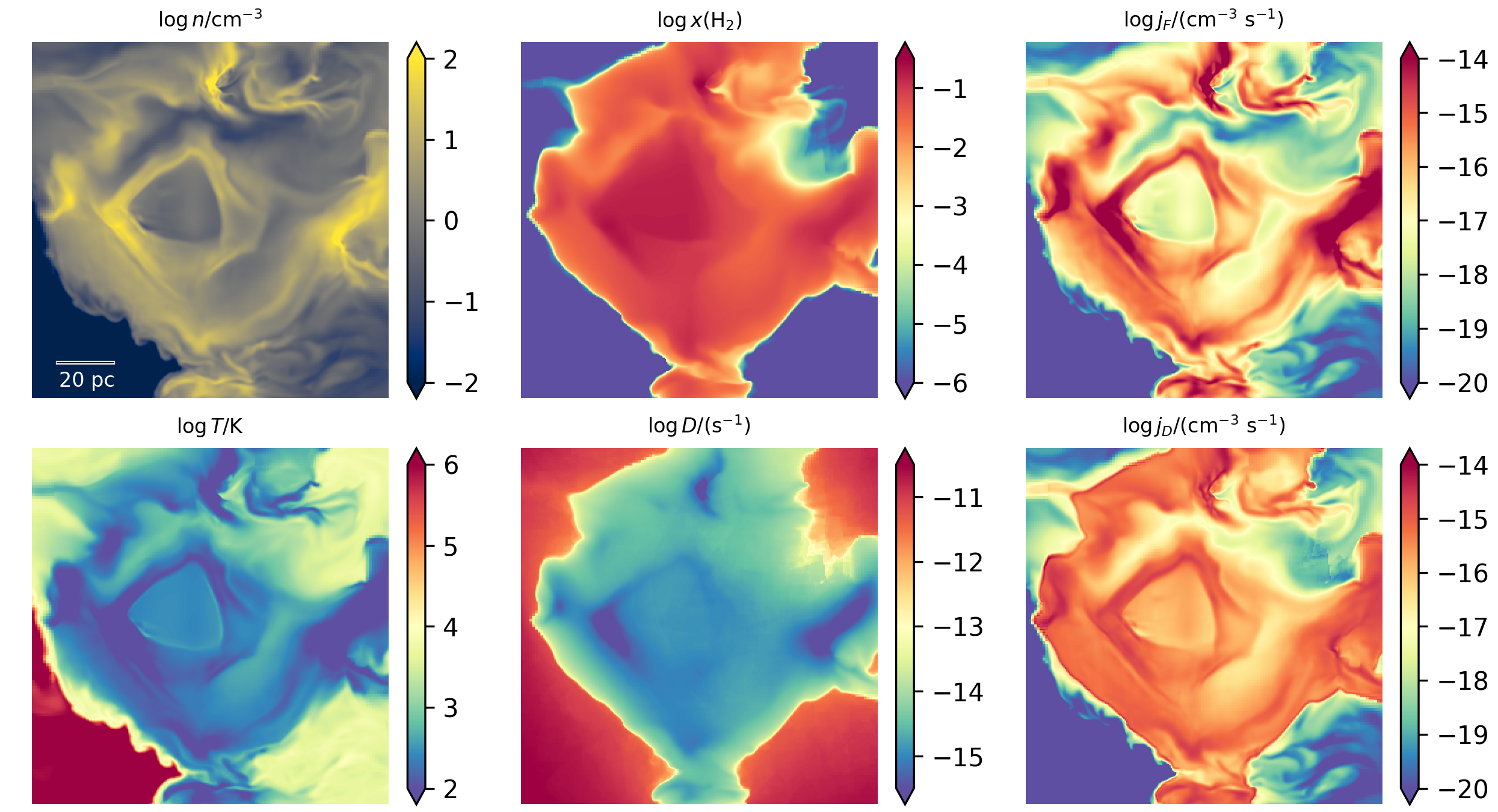}
    \caption{A 2D slice (xy plane) through the SILCC-Zoom simulation. The box size is $125 {\rm pc} \times 125 {\rm pc}$.
    The different panels show the gas density, the gas temperature, the H$_2$ abundance ($x({\rm H_2}) \equiv n({\rm H_2})/n$), the H$_2$ photodissociation rate (Eq.~\ref{eq: D}), and the volumetric H$_2$ formation and photodissociation rates (Eq.~\ref{eq: jf jd}).
     For an interactive figure, see \href{https://sbialy.wixsite.com/astro/visuals}{sbialy.wixsite.com/astro/visuals}.}
    \label{fig: slices} 
\end{figure*}

\begin{table}
\caption{Simulation Properties and Chemical State Overview}
\begin{center}
\begin{tabular}{l l}    
    \toprule
        Quantity & value \\ \midrule
        Simulation side length & $125$ pc \\
        Resolution & $512^3$ \\
        Simulation time (fiducial snapshot) & 3 Myr \\
        Mean density & $3.3$ cm$^{-3}$ \\
        Radiation field strength ($\chi$) & 1 \\
        Total gas mass & $2.2 \times 10^5 \ M_{\odot}$    \\   
         & \\   \midrule  
        Fraction of gas out of CSS & 41\% (by volume) \\
          & 74\% (by mass) \\ 
        \bottomrule
    \end{tabular}
    \end{center}
    The table is divided into two sections. The top section summarizes the simulation setup, including key physical parameters. The bottom section provides an overview of the chemical state of the gas, with chemical steady state (CSS) defined as cells satisfying $0.5<(j_F/j_D)<2$.
    For summary statistics of additional time snapshots, see Table \ref{tab:rates_comparison}
    \label{tab:table sim properties}
\end{table}

\section{Numerical Method}
\label{sec: Numerical Model}

\subsection{Hydro Simulations}
\label{sub:num}

We generate synthetic maps of H$_2$ line emission, formation rate, and dissociation rate using the SILCC-Zoom simulations \citep{Seifried2017}. These are high-resolution zoom-in runs derived from the SILCC simulation suite \citep{Walch2015b, Girichidis2015a}.
The SILCC simulation suite is a set of magnetohydrodynamical models of the 
chemical and thermal state of the ISM and the
formation of stars in realistic galactic environments.
The SILCC simulations have a stratified-box geometry.  They include self-gravity as well as a background potential for the
Galactic disk.  
Each simulation follows the thermal evolution of
the gas and dust, including photoelectric heating by dust, cosmic-ray
%and X-ray
ionization heating, and radiative cooling through various atoms ions, and molecules.
The chemistry of the ISM is modeled using an ``on-the-fly" time-dependent network for hydrogen and carbon chemistry,
tracking the evolution of the chemical abundances of free electrons, O, H$^+$, H, H$_2$, C$^+$, and CO. 
The simulation models H-H$_2$ chemistry, encompassing H$2$ formation on dust, photodissociation, photoionization, and cosmic-ray ionization. It also accounts for the attenuation of non-ionizing LW radiation through dust absorption ($f_{\rm dust}$) and H$_2$ self-shielding ($f_{\rm H_2, shield}$). These attenuation factors, as represented in Eq.~\ref{eq: D}, are calculated using the TreeRay algorithm \citep{Wunsch2018}, which considers radiation propagating along multiple directions for each cell.

The SILCC-Zoom simulations enhance the base SILCC framework by implementing adaptive mesh refinement, where regions of impending molecular cloud formation are resolved down to 0.12 pc scales, compared to the base 4 pc resolution used for driving turbulence through supernova injection.
The specific zoom-in run used herein is the ``MC1-MHD" simulation, a simulation which contains an initial magnetic field with a strength of 3~$\mu$G,
described first in detail in \citet[][]{Seifried2019,Seifried2020}.
%\citet{Seifried2017} \citep[see also][]{Haid2019, Seifried2020}. 

In this paper, we evaluate the reliability of using H$_2$ emission lines in combination with gas column densities to trace H$_2$ formation and dissociation rates. Specifically, we investigate whether the method described in \S \ref{sub: H2 formation dissociation - obs} can accurately identify regions that deviate from CSS. For our analysis, we focus on a snapshot taken at 3 Myr after initiating the zoom-in procedure. No star formation was considered in the simulations, which allows us to focus on the formation process of the cloud itself.
%before the onset of star formation. 
At this point, the effects of non-CSS conditions are most pronounced.
While this represents a simplified scenario compared to realistic star-forming clouds, it allows us to establish and validate our methodology in a controlled setting. The extension to more complex environments, including actively star-forming regions and the effects of supernova/stellar feedback, is explored in our companion paper \citep{Burkhart2024}. 
We note that our analysis is most applicable to clouds with column densities $N \lesssim 2 \times 10^{22}$ cm$^{-2}$, where cosmic-ray ionization and excitation remains subdominant (see \S \ref{sub: discuasion - CRs} and Appendix \ref{app: CRs}).

For the analysis presented in this paper we map the zoom-in region over an extent of (125~pc)$^{3}$ to a uniformly resolved grid with a resolution of 0.244~pc, i.e.~512$^{3}$ cells. The general properties of the extracted volume are given in Table~\ref{tab:table sim properties}.
We consider additional time snapshots and quantify the robustness of our analysis in Appendix \ref{app: snapshots stats}.

\subsection{Synthetic Observations}
\label{sub: synthtic obs}

The SILCC-Zoom simulations output the atomic, molecular, and total H nuclei volume densities, $n({\rm H})$, $n({\rm H_2})$, and $n$, respectively, the gas and dust temperatures, $T$ and $T_d$, and the LW radiation attenuation factors, $f_{\rm H_2, shield}$ and $f_{\rm dust}$, on a cell-by-cell basis. 
With these outputs, we 
derive the H$_2$ line emission map ($\mathcal{I}_{\rm tot}$) as follows:
\begin{enumerate}
\item We consider the observer-cloud LOS extending along the $x$-axis direction.
\item For each cell at ``sky" position $y,z$, and depth $x$ in the simulation, the hydrogen nucleus column density from that cell to the observer is $N(x,y,z) = \int_0^x n(x', y, z) \mathrm{d}x'$, and the corresponding optical depth is $\tau(x,y,z) = \sigma N(x,y,z)$.
% where we adopt $\sigma=1.9 \times 10^{-21}$ cm$^2$ \citep{Sternberg2014}.  
\item For each cell in the simulation, $(x,y,z)$, we have the optical depth $\tau(x,y,z)$, the local photodissociation rate $D$ (Eq.~(\ref{eq: D}) with $\chi=1$), and the local excitation rate $P=D/p_{\rm diss} \approx D/0.15$ (Eq.~\ref{eq: p_diss}).
Using these quantities, we calculate the 2D map of $\mathcal{I}_{\rm tot}$ for all LOSs $(y,z)$, using Eq.~(\ref{app eq: Itot 2}), where the integration is along $x$.
\item We repeat steps 1--3 for the two other LOS orientations, LOSs along the $y$- and $z$-axes.
\end{enumerate}
At the conclusion of steps 1--4, we obtain maps of the H$_2$ emission line intensity, $\mathcal{I}_{\rm tot}$, for three LOS orientations. These maps are presented in \S \ref{sub; results 2D maps}. It is important to note that our synthetic observations are idealized, as they do not account for the limitations of observational instruments. 
% Instead, they assume resolution and sensitivity constrained only by the parameters of the SILCC simulation itself.

Utilizing Eq.~(\ref{eq: D-Itot}), we convert $\mathcal{I}_{\rm tot}$ to $\dot{\Sigma}_{D}^{\rm (obs)}$.
In what follows (\S \ref{sub; results 2D maps}-\ref{sub: PDFs of JF and JD}) we compare these observationally-derived rates, $\dot{\Sigma}_{D}^{\rm (obs)}$ with the true rates, $\dot{\Sigma}_{D}^{\rm (true)}$ as given by directly integrating the volumetric formation and dissociation rates in the simulation (Eq.~\ref{eq: JF JD}).
This comparison addresses how well the total H$_2$ line intensity traces the H$_2$ photodissociation rate.
Similarly, we derive $\dot{\Sigma}_{F}^{\rm (obs)}$ and compare it with the true formation rate given by the simulation  $\dot{\Sigma}_{F}^{\rm (true)}$.
To obtain $\dot{\Sigma}_{F}^{\rm (obs)}$, we utilize the HI column density, $N({\rm H})$, and the total (HI+H$_2$) gas column density, $N$, obtained from the simulation data.
We then use Eq.~(\ref{eq: JF observation 2}) to derive $\dot{\Sigma}_{F}^{\rm (obs)}$.
As is the case for $\dot{\Sigma}_{D}^{\rm (obs)}$, the formation rate $\dot{\Sigma}_{F}^{\rm (obs)}$ is an approximation for the true rate, as it relies on an average relation (Eq.~\ref{eq: JF observation 2}) that uses integrated quantities (which can be observed) as inputs, as opposed to the real, cell-by-cell volumetric formation rates.

Our analysis assumes the availability of reliable total gas column density maps. In practice, deriving such maps from observations requires careful consideration of multiple tracers: HI 21 cm emission for atomic gas, CO lines with appropriate $X_{\rm CO}$ conversion factors \citep{Bolatto2013}, and dust continuum emission assuming dust-to-gas ratios and emissivity laws \citep{PlanckCollaboration2011}. These conversions can introduce uncertainties of factors of 2-3 in the derived column densities. While these observational challenges are crucial for applying our method to real data, addressing them is beyond the scope of this paper. Instead, in this paper, we focus on the subsequent question: given maps of HI and total gas column density derived from observations, how well can we constrain H$_2$ formation and dissociation rates, and assess whether the gas is in CSS?

In \S \ref{sub; results 2D maps}-\ref{sub: PDFs of JF and JD}, we compare $\dot{\Sigma}_{F}^{\rm (obs)}$ with $\dot{\Sigma}_{F}^{\rm (true)}$ by directly integrating the volumetric H$_2$ formation rate, cell-by-cell, along the LOS (Eq.~\ref{eq: JF JD}).

\begin{figure}[t]
    \centering
    \includegraphics[width=0.5\textwidth]{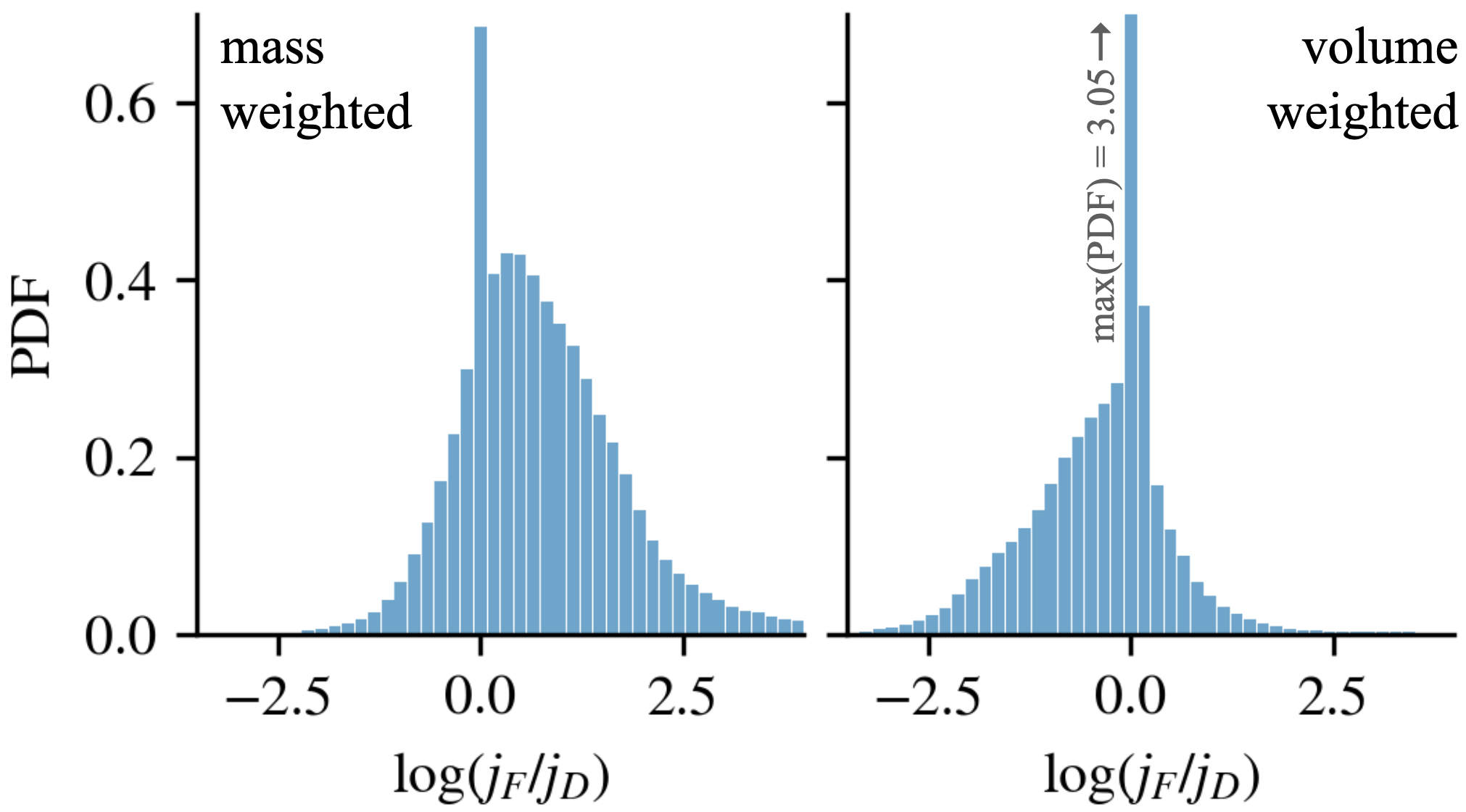}
    \caption{The 1D PDF of $\log j_F/j_D$, weighted by mass (left) and volume (right). Each PDF is composed of a population of cells that are in CSS (the peak at $\log j_F/j_D=0$), and a wide and prominent distribution of gas that is out-of-CSS  (see also Table \ref{tab:table sim properties}). 
    }
    \label{fig: vol quantities 1D PDFs} 
\end{figure}

\begin{figure*}
    \centering
    \includegraphics[width=1.0\textwidth]{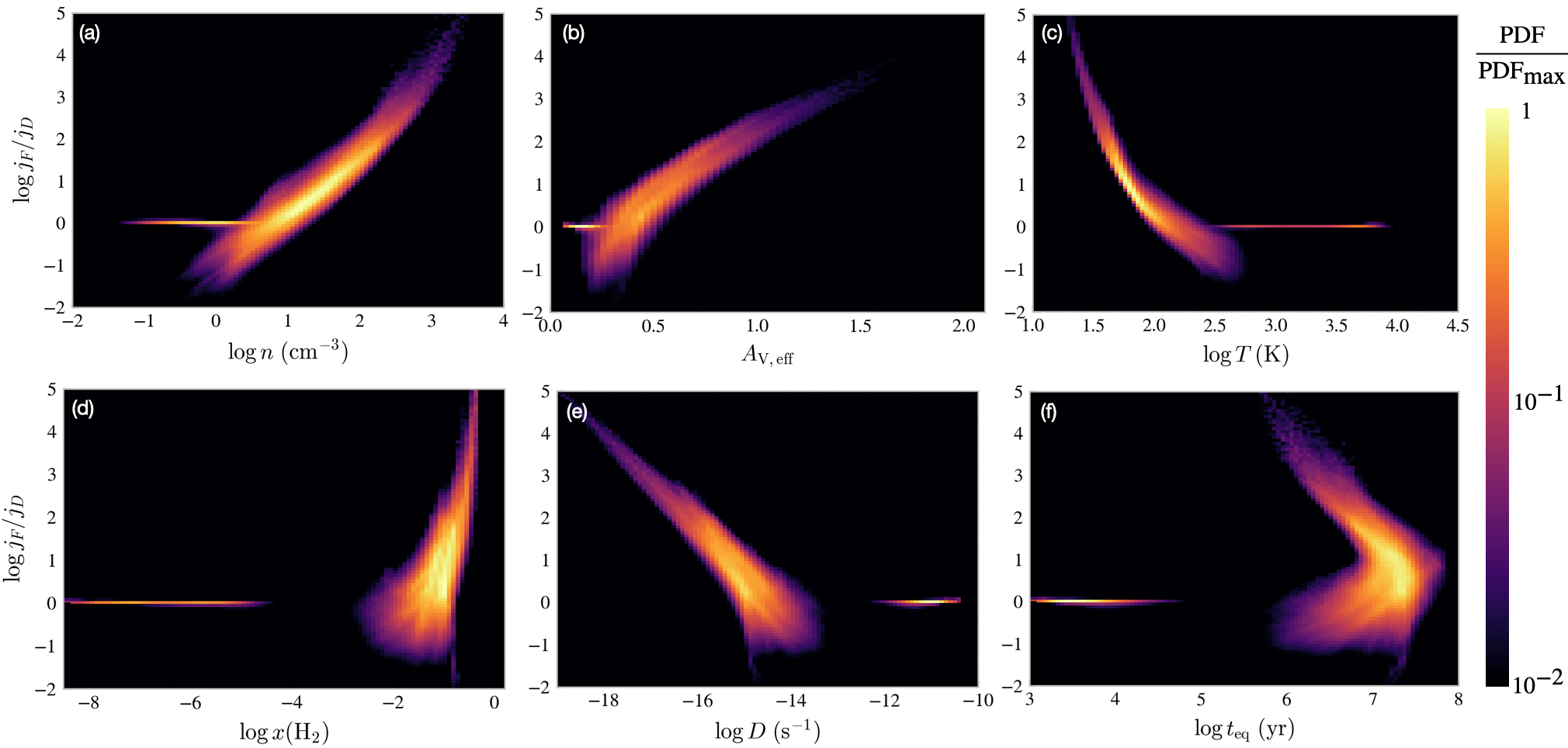}
    \caption{Mutual 2D mass-weighted PDFs of volumetric quantities in the simulation box.
    We show 2D PDFs of $\log j_F/j_D$ versus (a) gas density; (b) effective dust extinction; (c) gas temperature; (d) H$_2$ fraction; (e) local (attenuated) photodissociation rate; (f) chemical timescale (see text for details).
    These PDFs reveal two distinct populations: 
    [1] A warm, diffuse and H$_2$-poor gas, located in low-shielded regions, that has achieved CSS (i.e., the narrow horizontal strips at $\log j_F/j_D=0$) and [2] Cold, dense, H$_2$-rich, well-shielded gas, that is out of CSS and exhibits a wide $j_F/j_D$ distribution.
    }
    \label{fig: vol quantities} 
\end{figure*}

\section{Results}
\label{sec: results}

\subsection{Volumetric quantities}
\label{sub: resuluts form-dest volumetric}

Before presenting the integrated H$_2$ formation/dissociation rates, we begin by exploring key volumetric quantities. This provides intuition regarding the conditions in the simulation box.
On all figures, $\log$ denotes the logarithm base 10.

Fig.~\ref{fig: slices} illustrates the cloud's density structure and its chemical state.
It shows a 2D slice parallel to the $xy$ plane sliced at the middle of the $z$ axis. The six panels correspond to various fields: the gas density $n$, the gas temperature $T$, the H$_2$ abundance $x({\rm H_2}) \equiv n({\rm H_2})/n$, the local H$_2$ (shielded) photodissociation rate $D$ (Eq.~\ref{eq: D}),
and the volumetric H$_2$ formation and photodissociation rates, $j_F$ and $j_D$ (Eq.~\ref{eq: jf jd}). 
The gas is highly inhomogeneous, with density and temperature spanning large ranges.
Due to absorption of the FUV radiation, the cloud interior is mostly cold, with $T \lesssim 100$~K.
This attenuation of the radiation intensity with cloud depth is also evident in the maps of $x({\rm H_2})$ and $D$ where we see a sharp decrease in the photodissociation rate, and a sharp increase in the H$_2$ abundance from cloud edge to the cloud interior.
The clumpy density structure of the cloud, as well as time-dependent chemical effects, result in inhomogenous structures of $j_F$ and $j_D$ where often the two are not in balance.

In Fig.~\ref{fig: vol quantities 1D PDFs} we present the probability density function (PDF) of  $j_F/j_D$ (Eq.~\ref{eq: jf jd}), weighted by mass (left) and by volume (right).
The PDFs are composed of two components: a sharp peak at $\log j_F/j_D=0$ which arises from gas cells that are in CSS, and a broad component spanning a large range of out-of-CSS gas.

\begin{figure*}[t]
    \centering
    \includegraphics[width=1\textwidth]{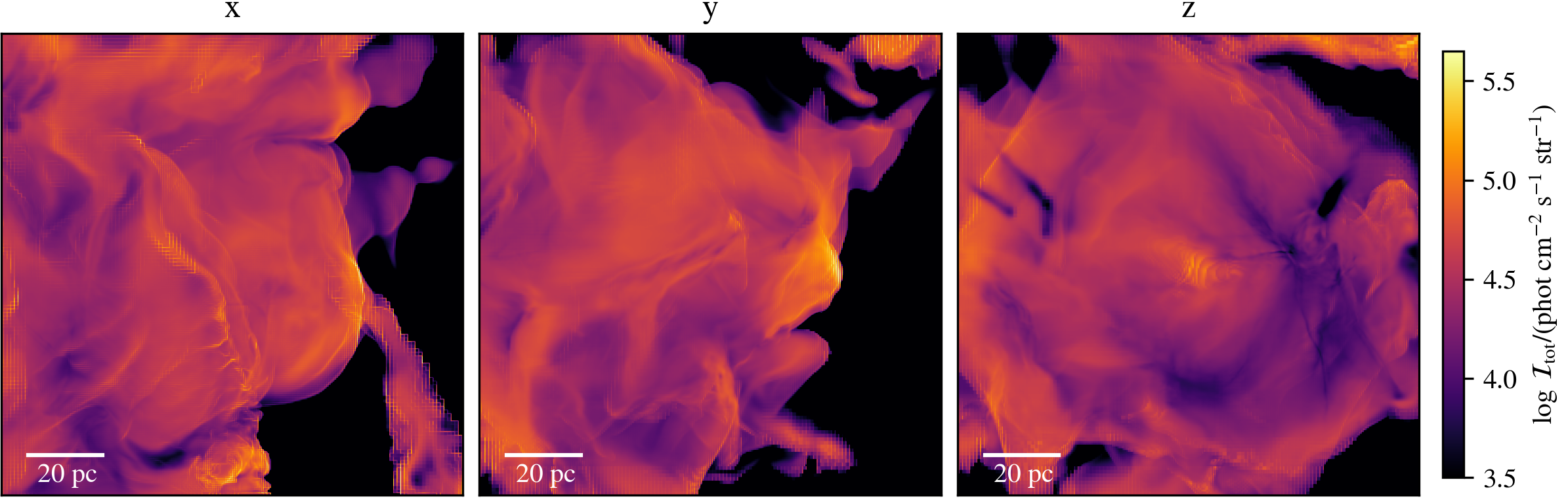}
    \caption{H$_2$ line emission maps. The panels show the H$_2$  emission line intensity (summed over all emission lines), for three orientations of the observer's LOS (see \S \ref{sub: synthtic obs} for details).}
    \label{fig: I line} 
\end{figure*}

To get insight onto these two populations, in Fig.~\ref{fig: vol quantities} we plot the joint distribution (mass-weighted) of $j_F/j_D$ versus  various volumetric quantities: 
    (a) the gas number density $n$; 
    (b) the effective dust extinction $A_{\rm V, eff}$ \footnote{$A_{\rm V, eff}$ is defined as a weighted mean of $A_V$ from the cell to the cloud edge for different rays, and is related to the dust shielding factor through
    $f_{\rm dust} = \mathrm{e}^{-3.5 A_{\rm V, eff}}$ \citep{Seifried2020}. \label{footnote Aveff} }
    (c) the gas temperature, $T$;
    (d) the H$_2$ abundance $x({\rm H_2})$;
    (e) the local H$_2$ photodissociation rate, $D$ (Eq.~\ref{eq: D});
    (f) the  timescale required to reach CSS, $t_{\rm chem}$ (Eq.~\ref{eq: t chem}).
Looking at the 2D distributions in Fig.~\ref{fig: vol quantities}, we see these two populations.
In the upper panels we see that the CSS population corresponds to gas that is relatively diffuse and warm, with $n \sim 0.1-3$ cm$^{-3}$ and $T \approx 300-6000$ K, and has low visual extinctions, $A_{\rm V, eff} \lesssim 0.25$. This gas is located closer to the cloud boundaries and is exposed to relatively strong FUV radiation.
Indeed, from the $D$ and $x_{\rm H_2}$ PDFs, we see that the CSS gas experiences high dissociation rates $D \approx 10^{-12}-10^{-10}$ s$^{-1}$ (i.e., little attenuation), and is predominantly atomic with $x({\rm H_2}) \lesssim 10^{-5}$. 
The timescale to achieve CSS for such conditions is the H$_2$ photodissociation time and is very short, $t_{\rm chem} \simeq 1/D \sim (0.03-3) \times 10^4$ yrs, and thus the gas is in CSS.

The gas that is in CSS occupies a significant volume fraction, however, it includes only a small fraction of the MC mass.
Most of the mass is found in out-of-CSS gas \citep[see also][]{Seifried2021}. This population shows up in Fig.~\ref{fig: vol quantities} as the wide distribution of pixels extending from $\log j_F/j_D \sim -2$ to $\sim 4$. This gas is typically denser and colder, $n \sim 1-10^3$ cm$^{-3}$, $T \approx 20-300$ K, and is located in inner cloud regions with $A_{\rm V, eff}=0.25-1.5$. This gas is exposed to low FUV intensities ($D \lesssim 10^{-14}$ s$^{-1}$) and is H$_2$-rich ($x({\rm H_2}) \gtrsim 10^{-2}$). Under these conditions, the chemical timescale is long, $t_{\rm chem} \simeq 1/(2Rn) \approx 1\text{--}100~\text{Myr}$, and the gas has not had sufficient time to reach CSS.

Qualitatively, defining CSS as regions where the H$_2$ formation and dissociation rates differ by less than a factor of two (i.e., $0.5<(j_F/j_D)<2$), we find that only 26 \% of the simulation mass is in CSS (see Table \ref{tab:table sim properties}).
The volume fraction of cells that are in CSS is 59 \%.
The volume fraction is higher than the mass fraction because 
the CSS regions reside near cloud boundaries where the gas is typically more diffuse and thus occupies larger volumes.

Integrating $j_F$ and $j_D$ over the MC volume, we obtain a total H$_2$ formation rate
$\dot{M}_{\rm F}^{\rm (true)} = 8.6 \times 10^3 \ \mathrm{M_{\sun} \ Myr^{-1}}$, dissociation rate
$\dot{M}_{\rm D}^{\rm (true)} = 1.1 \times 10^3 \ \mathrm{M_{\sun} \ Myr^{-1}}$, 
and a net H$_2$ formation $\dot{M}_{\rm H_2}^{\rm (true)} = \dot{M}_F^{\rm (true)} - \dot{M}_D^{\rm (true)} = 7.5 \times 10^3 \ \mathrm{M_{\sun} \ Myr^{-1}}$.
This net positive formation rate indicates that if the cloud were to maintain these gas conditions for a sufficiently long time ($t > t_{\rm chem}$), a significant mass of H\textsc{i} would eventually convert to H$_2$. Consequently, the reduced H\textsc{i} fraction would cause $j_F$ to decrease (see Equation~\ref{eq: jf jd}) until it finally equilibrates with $j_D$, at which point the gas reaches CSS.
In Paper II of this series \citep{Burkhart2024}, we explore the evolution of H$_2$ formation and dissociation rates over the dynamical timescales of MCs.

\begin{figure}[t]
    \centering \includegraphics[width=0.5\textwidth]{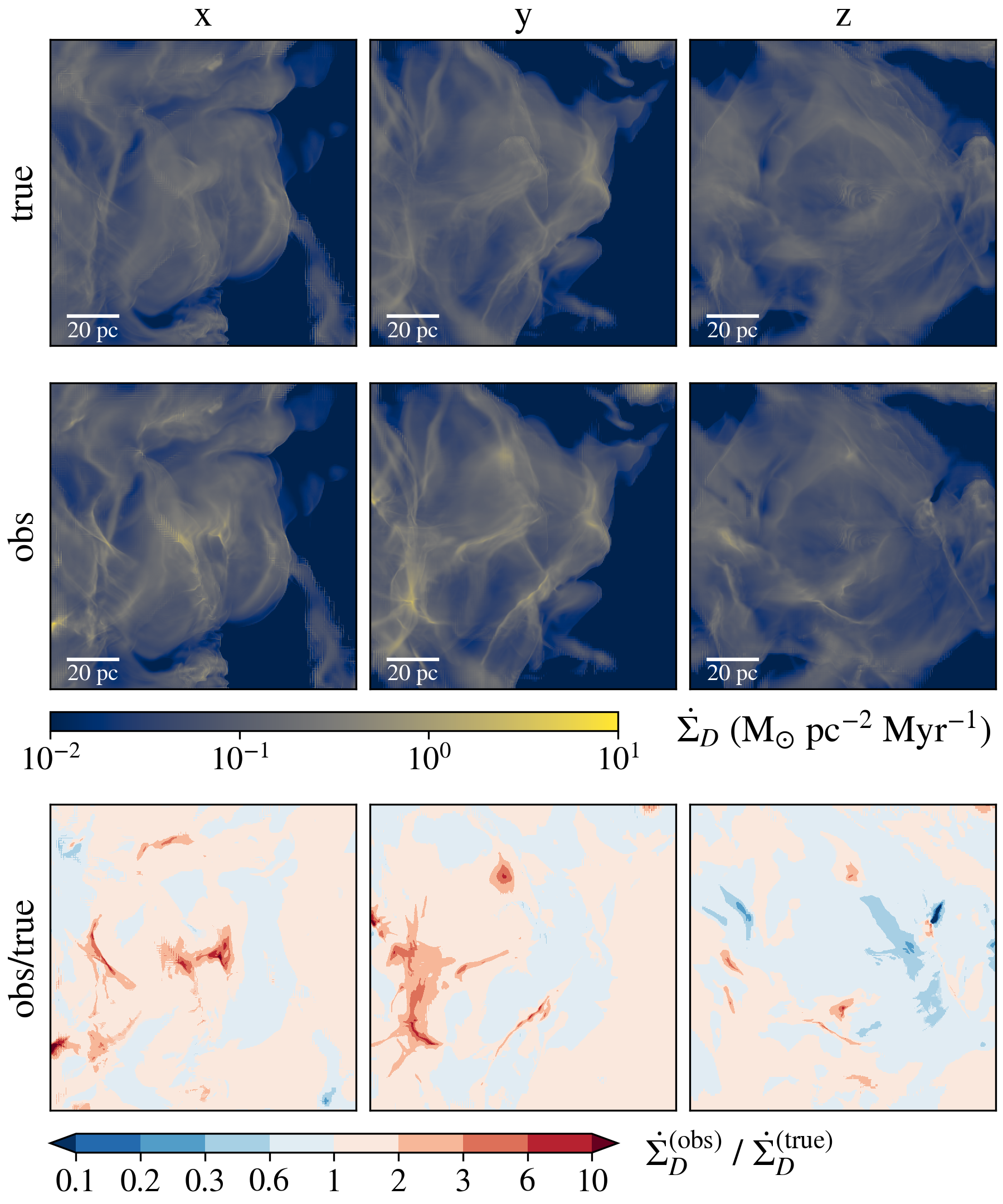}
    \caption{Integrated H$_2$  photodissociation rate maps. 
    {\bf Top}: maps of the {\it true} photodissociation rate, $\dot{\Sigma}_{D}^{\rm (true)}$, as calculated by integrating the volumetric photodissociation rate (Eq.~\ref{eq: JF JD}).
    {\bf Middle}: The observationally-derived rate, $\dot{\Sigma}_{D}^{\rm (obs)}$, as derived from the H$_2$ line emission map, $\mathcal{I}_{\rm tot}$ (Eq.~\ref{eq: D-Itot}; \S \ref{sub: synthtic obs}).
    {\bf 3$^{\rm rd}$ Bottom}: The ratio $\dot{\Sigma}_{D}^{\rm (obs)}/\dot{\Sigma}_{D}^{\rm (true)}$.}
    \label{fig: maps JD} 
\end{figure}

\begin{figure}[t]
    \centering
    \includegraphics[width=0.5\textwidth]{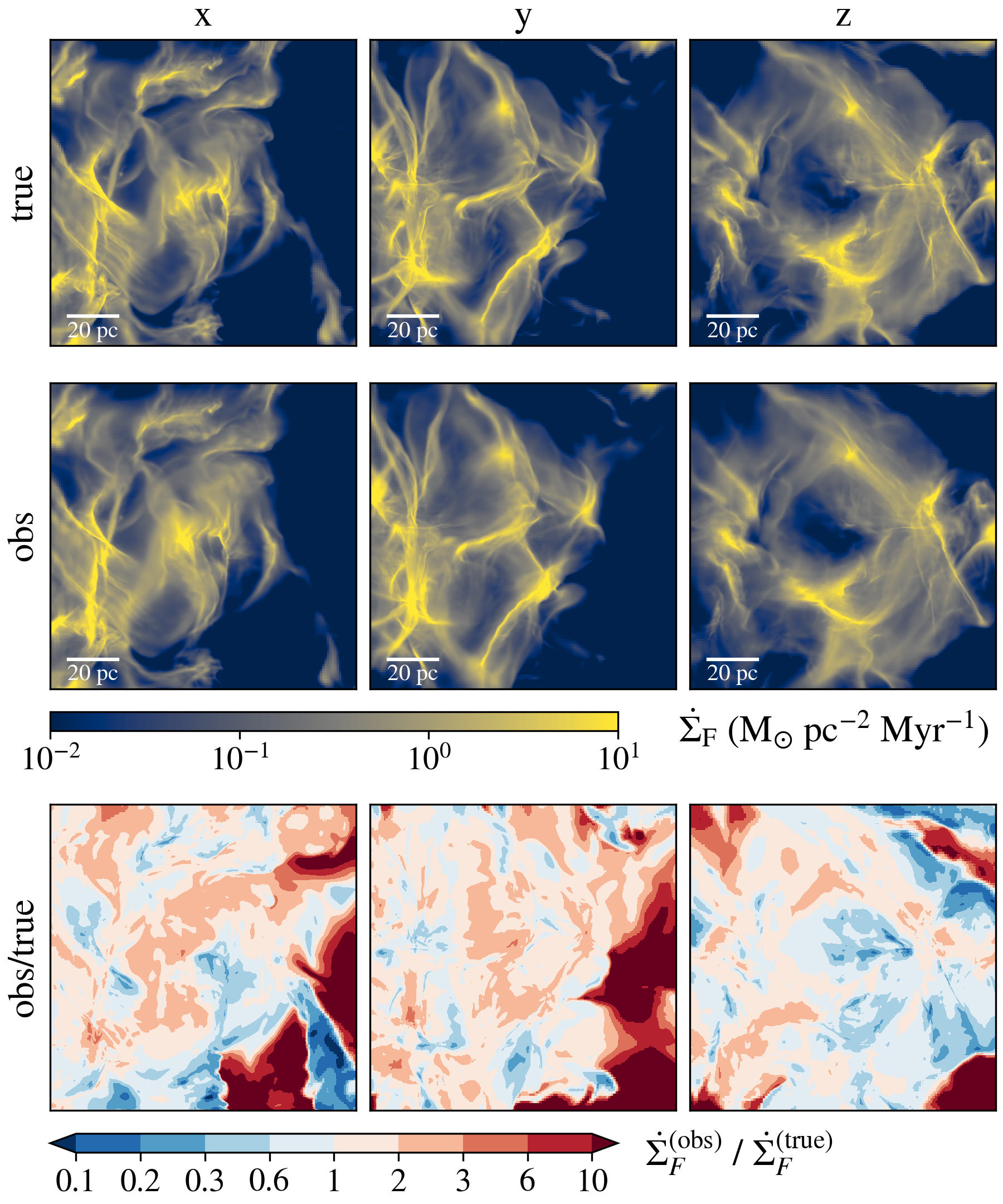}
    \caption{As Fig.~\ref{fig: maps JD}, but for the H$_2$ formation rate.}
    \label{fig: maps JF} 
\end{figure}

\begin{figure}
    \centering
    \includegraphics[width=0.5\textwidth]{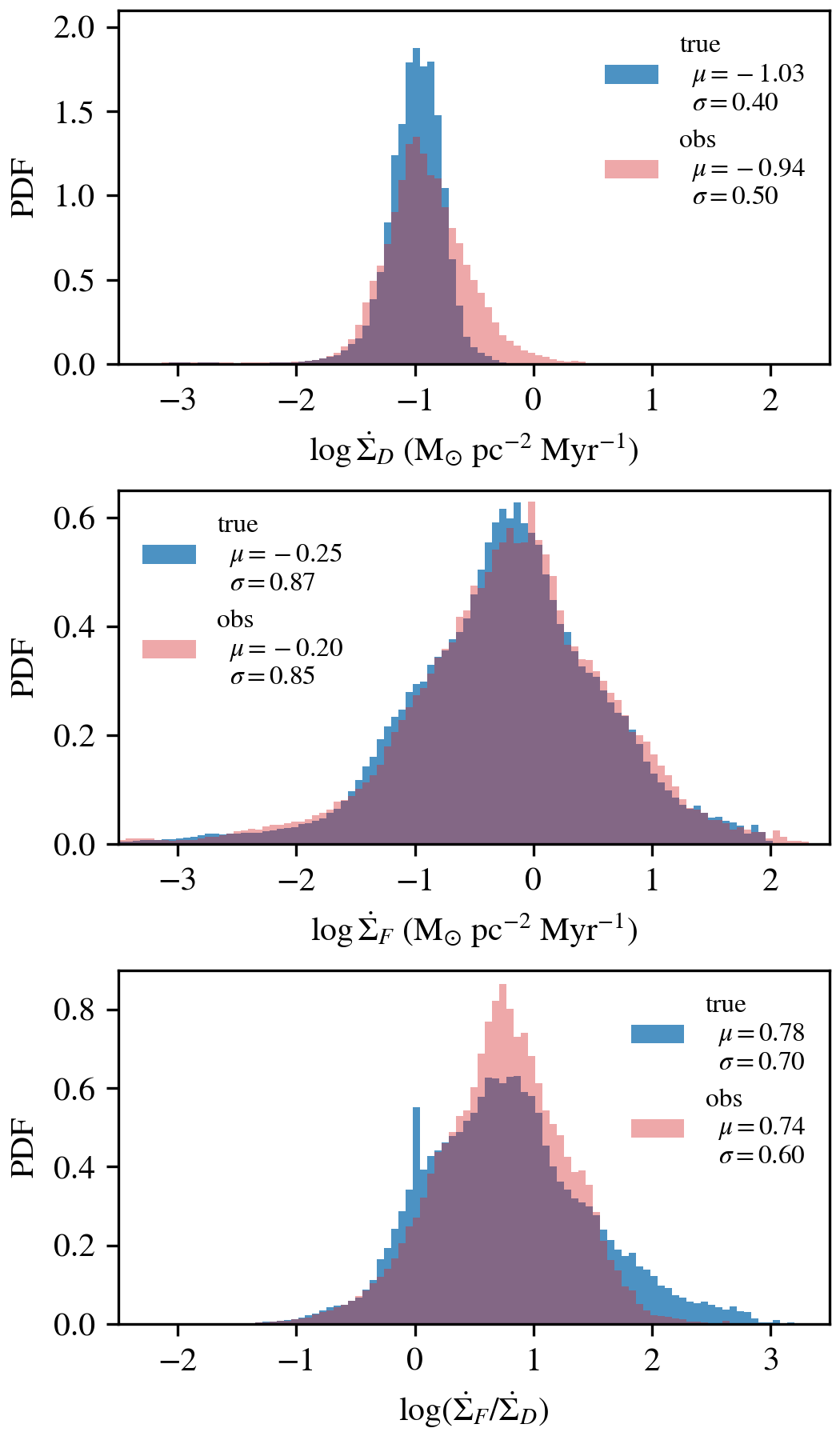}
    \caption{
    Mass-weighted rate PDFs. 
    The PDF of the H$_2$ photodissociation rate, $\log \dot{\Sigma}_{D}$ (top),   
    the H$_2$ formation rate, $\log \dot{\Sigma}_{F}$ (middle) and their ratio, $\log (\dot{\Sigma}_{F}/\dot{\Sigma}_{D})$ (bottom). For these PDFs we stacked data for the three LOS orientations, $x$, $y$, $z$.
    Each panel compares the observationally-derived (Eqs.~\ref{eq: D-Itot}, \ref{eq: JF observation 2}) and the true (Eq.~\ref{eq: JF JD}) rate PDFs . 
    The mean and standard deviations are indicated.
    }
    \label{fig:JF JD PDFs} 
\end{figure}
\begin{figure}
    \centering
    \includegraphics[width=0.5\textwidth]{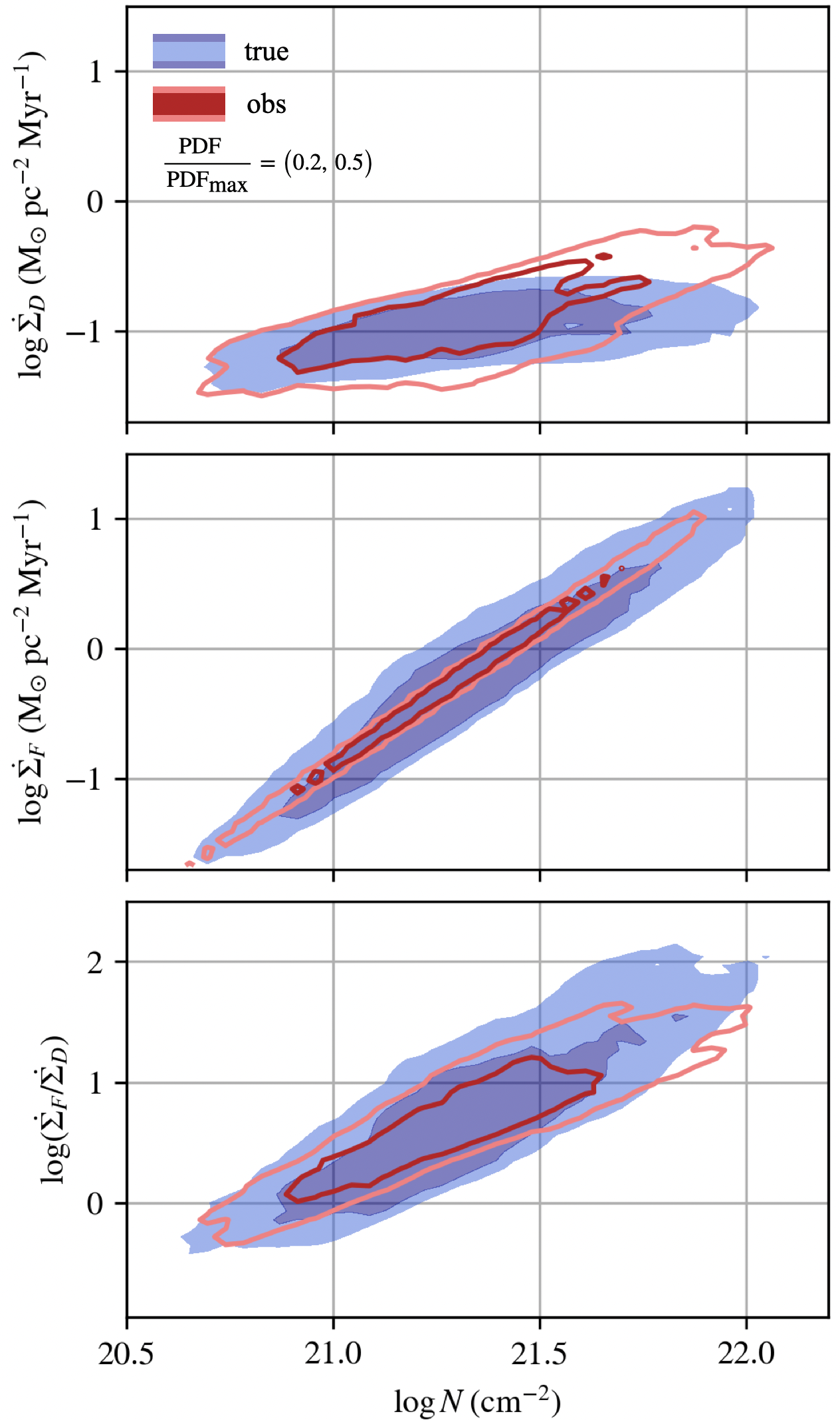}
    \caption{
    Mass-weighted mutual 2D PDFs of $\log N$, the total gas column density along the LOS,  versus
    $\log \dot{\Sigma}_{D}$ (top),   
    $\log \dot{\Sigma}_{F}$ (middle) and $\log (\dot{\Sigma}_{F}/\dot{\Sigma}_{D})$ (bottom).
    The red and blue PDFs correspond to the observationally-derived (Eqs.~\ref{eq: D-Itot}, \ref{eq: JF observation 2}) and the true (Eq.~\ref{eq: JF JD}) rate PDFs respectively. The light and dark red contours correspond to PDF/PDF$_{\rm max}=0.2$ and 0.5, respectively. Similarly, the light blue shows the level $0.2 \leq {\rm PDF/PDF_{\rm max}} < 0.5$ and the darker blue shows the PDF region of  ${\rm PDF/PDF_{\rm max}} \geq 0.5$.}
    \label{fig:JF JD vs N 2d PDFs} 
\end{figure}

\subsection{ 2D Maps }
\label{sub; results 2D maps}

Fig.~\ref{fig: I line} presents maps of the total H$_2$ line emission, $\mathcal{I}_{\rm tot}$, for three LOS orientations, generated following the procedure described in \S~\ref{sub: synthtic obs}. These maps simulate the observations an astronomer might obtain by directing an FUV spectrograph toward an MC and aggregating all observed H$_2$ line intensities.

Fig.~\ref{fig: maps JD} compares the 2D maps of the true H$_2$ photodissociation rate, $\dot{\Sigma}_{D}^{\rm (true)}$, and the observer-derived rate, $\dot{\Sigma}_{D}^{\rm (obs)}$. The true rate is calculated by integrating the volumetric photodissociation rate cell-by-cell in the simulation (Eq.~\ref{eq: JF JD}), while the observer-derived rate is obtained from the H$_2$ line emission maps (Eq.~\ref{eq: D-Itot}; see also \S\ref{sub: synthtic obs}). For the three LOS orientations, the observer-derived maps effectively recover the true H$_2$ dissociation maps over a large dynamic range of photodissociation rates, from $\dot{\Sigma}_{D}=10^{-2}$ to $10 \ M_{\odot}$ pc$^{-2}$ Myr$^{-1}$.  
As discussed in \S\ref{sub: H2 dissociation - obs}, the observationally-derived rates and true rates are not identical because an observer does not have access to the true value of the attenuation factor and must rely on integrated quantities (i.e., see Appendix \ref{app: Sigma-Itot}; Eq.~\ref{app eq: g_dust approx} vs. Eq.~\ref{app eq: g_dust}).

Fig.~\ref{fig: maps JF} shows the true and observationally-derived H$_2$ formation rate maps. Here again, we observe that the overall structure of the observationally-derived H$_2$ formation rate map qualitatively agrees with the true formation rate map, spanning from weakly H$_2$-forming regions with $\dot{\Sigma}_{F}=0.01 \ M_{\odot}$ pc$^{-2}$ Myr$^{-1}$ to highly efficient H$_2$-forming regions with $\dot{\Sigma}_{F}=10 \ M_{\odot}$ pc$^{-2}$ Myr$^{-1}$.

Integrating $\dot{\Sigma}_{D}$ and $\dot{\Sigma}_{F}$ over the area, we obtain the total H$_2$ formation and dissociation mass rates. Using the observationally-derived surface densities $\dot{\Sigma}_{D}^{\rm (obs)}$ and $\dot{\Sigma}_{F}^{\rm (obs)}$ (Eqs.~\ref{eq: D-Itot} and \ref{eq: JF observation 2}), we obtain $\dot{M}_F^{\rm (obs)}=(9.3, 10.8, 7.3) \times 10^3$ M$_{\odot}$ Myr$^{-1}$ and $\dot{M}_D^{\rm (obs)}=(1.4, 1.5, 1.1) \times 10^3$ M$_{\odot}$ Myr$^{-1}$ for the $x$, $y$, and $z$ LOS orientations, respectively.
We compare these measurements with the true mass rates (independent of orientation): $\dot{M}_F^{\rm (true)}=8.6 \times 10^3$ M$_{\odot}$ Myr$^{-1}$ and $\dot{M}_D^{\rm (true)}=1.1 \times 10^3$ M$_{\odot}$ Myr$^{-1}$. The true rates are obtained by integrating $\dot{\Sigma}_{D}^{\rm (true)}$ and $\dot{\Sigma}_{F}^{\rm (true)}$ (Eq.~\ref{eq: JF JD}), or equivalently, by volumetrically integrating $j_F$ and $j_D$ (Eq.~\ref{eq: jf jd}).

To assess the robustness of these results, we extend this analysis to different simulation snapshots ($t=2$, $3$, $4$, and $5$ Myr) in Appendix~\ref{app: snapshots stats}, comparing the observationally-derived rates with the true values. Across all realizations (4 snapshots $\times$ 3 LOS orientations), the observationally-derived rates agree with the true rates to within a mean relative difference of 27\% for H$_2$ formation and 31\% for H$_2$ dissociation (see Table~\ref{tab:rates_comparison} for detailed statistics).

\subsection{PDFs of $\dot{\Sigma}_{D}$ and $\dot{\Sigma}_{F}$}
\label{sub: PDFs of JF and JD}

Fig.~\ref{fig:JF JD PDFs} shows the mass-weighted PDFs of $\dot{\Sigma}_{D}$, $\dot{\Sigma}_F$ and $\dot{\Sigma}_{F}/\dot{\Sigma}_{D}$.
For these PDFs, we stack the data for the three LOS orientations, $x$, $y$, $z$.
In each panel, the blue histograms correspond to the true rates, and the red to the observationally-derived rates. 
Qualitatively, the observational PDFs provide a good approximation to the true PDFs, recovering the general shape, average position and PDF dispersion.
Quantitatively, we find that there are (small) statistical differences.
For example, the mass-weighted average and standard deviation of $\log \dot{\Sigma}_{D}$ are $(\mu, \sigma)=(-1.03, 0.40)$  for the ``true" PDF, and $(-0.94, 0.50)$ for the ``observed" PDF.
For $\log \dot{\Sigma}_{F}$ we find $(\mu, \sigma)=(-0.25, 0.87)$  for the ``true" PDF, and
$(-0.20, 0.85)$ for the ``observed" PDF.

Comparing the PDFs in the top vs the middle panels, we see that the formation-rate PDFs have significantly larger dispersion compared to the dissociation-rate PDFs.
This high dispersion in $\dot{\Sigma}_{F}$ is driven by the strong density fluctuations of the cloud, and the fact that for a significant fraction of the cloud mass the gas has not yet reached CSS (see \S \ref{sub: resuluts form-dest volumetric}).

This is further demonstrated in Fig.~\ref{fig:JF JD vs N 2d PDFs}. 
in which we show the joint (2D) mass-weighted PDFs of $\dot{\Sigma}_{F}$, $\dot{\Sigma}_{D}$, and $\dot{\Sigma}_{F}/\dot{\Sigma}_{D}$, versus the gas column density $N$. 
Both formation and dissociation rates systematically increase with $N$.
However, the formation rate has a steeper slope. Consequently, at large column densities, the H$_2$ formation rate surpasses the photodissociation rate. As discussed in \S~\ref{sub: resuluts form-dest volumetric}, the cloud regions contributing to this excess formation are areas with high volume density, typically embedded deep in the cloud, which are in the process of converting H\textsc{i} into H$_2$ and have not yet reached CSS. If the MC were to maintain its physical conditions for a sufficiently long time ($\gg t_{\rm chem}$), these regions would eventually convert most of their H\textsc{i} into H$_2$, reducing the formation rate until it balances with the dissociation rate. Such balance has already been achieved (approximately) in the more diffuse cloud regions, at $N \lesssim 10^{21}~\mathrm{cm}^{-2}$.

Fig.~\ref{fig:JF JD vs N 2d PDFs} also highlights cloud regions where our method performs well statistically and where it is less accurate. For H$_2$ dissociation, in less dense cloud regions ($N \lesssim 2 \times 10^{21}~\mathrm{cm}^{-2}$), the observationally derived rates closely match the true rates. In denser regions, however, the observationally derived rates generally overestimate the true rate. At $N = 3 \times 10^{21}~\mathrm{cm}^{-2}$, this overestimation is typically a factor of 1.4 (comparing the means of the two PDFs), whereas at $N=10^{22}~\mathrm{cm}^{-2}$, the observational PDF's performance further declines, with the overestimation increasing to a factor of 2.2. As we discuss in \S~\ref{sub: discuasion - CRs}, our method is not reliable at large columns in any case, as CR dissociation may become significant in such environments. For H$_2$ formation, the observationally derived PDF statistically recovers the true PDF over the entire range of $N$. However, at any given $N$, it exhibits a systematically lower dispersion compared to the true rate PDF.

These results demonstrate both the potential and limitations of our observational method for deriving H$_2$ formation and dissociation rates across various cloud conditions, setting the stage for a more detailed examination of its implications and applicability, which we discuss in the following section.

\section{Discussion}
\label{sec:dis}
As we showed in this paper, large fractions of clouds' masses may be far from CSS, with H$_2$ formation and dissociation not balancing each other.
We demonstrated that observations of H$_2$ emission lines, in combination with gas column densities, may be used to constrain the H$_2$ formation and dissociation rates and to assess whether the gas along the observation LOS is in CSS.
It is important to note that our method provides LOS integrated rates. In real observations, multiple cloud structures may overlap along the same line of sight. In such cases, our method can reveal if some portion of the gas is out of CSS, but cannot resolve which specific clouds are responsible. This limitation suggests focusing initial applications on well-characterized single-cloud sightlines, such as high-latitude Galactic clouds.

\subsection{Generalization to IR line emission}
Since FUV line emission from H$_2$ excitation is followed by rovibrational cascades within the ground electronic state, it is accompanied by IR line emission \citep{VanDishoeck1986, Black1987, Sternberg1988, Sternberg1989}. This H$_2$ line emission has been observed in a variety of sources \citep[e.g.,][]{Gatley1987, Puxley1988, Dinerstein1988, Tanaka1989, Tanaka1991, Kaplan2021}. These IR emission lines can also be used to estimate the integrated H$_2$ photodissociation rate.

While IR line emission can be used to derive the H$_2$ photodissociation rate, a complication arises because these transitions originate from ro-vibrationally excited H$_2$ states, which can be excited by various mechanisms beyond photo-excitation, including collisional excitation in warm gas regions (e.g., shocks), secondary electrons produced by penetrating cosmic rays or X-rays, and chemical pumping during H$_2$ formation (Fig.~1 in \citealt{Bialy2020}). Consequently, deducing the H$_2$ photodissociation rate ($\dot{\Sigma}_{D}$) requires disentangling these various excitation processes. In contrast, FUV lines solely originate from photo-excitation, providing a more direct relation to the H$_2$ photodissociation rate, although IR line emission has the advantage of being less affected by dust extinction, allowing probing of deeper cloud regions. Ideally, both FUV and IR lines should be used in tandem to (a) verify the consistency of the H$_2$ dissociation rate derived from both methods, (b) assess the impact of dust extinction on H$_2$ line emission, and (c) determine the roles of the various excitation mechanisms.

Assuming the subtraction of H$_2$ line emission due to the alternative excitation processes, we 
may write a relation for the H$_2$ photodissociation rate in terms of the IR line emission.
Following the derivation of Eq.~(\ref{app eq: D-Itot v0}) (see Appendix \ref{app: Sigma-Itot}) we obtain:
\begin{align}
\label{eq: D-Itot IR}
    \dot{\Sigma}_{D}^{\rm (obs)} &= \frac{4 \pi p_{\rm diss} \bar{m} }{(1-p_{\rm diss}) \ \mathcal{N}^{\rm IR}} \   \mathcal{I}_{\rm tot}^{\rm IR} \\ \nonumber 
    &= 
    4.4 \times 10^{-2} \  \mathcal{I}_{5}^{\rm IR} \ {\rm M_{\odot}  \ pc^{-2} \ Myr^{-1}} \ ,
\end{align}
where $\mathcal{I}_{\rm tot}^{\rm IR}$ is the IR total line intensity, and $\mathcal{I}_{5} \equiv \mathcal{I}_{\rm tot}^{\rm IR}/(10^{5}$ photons cm$^{-2}$ s$^{-1}$ str$^{-1}$).
In Eq.~(\ref{eq: D-Itot IR}) we do not include the dust attenuation $\beta_{\rm dust}$ factor (see Appendix \ref{app: Sigma-Itot}, Eqs.~\ref{app eq: g_dust}, \ref{app eq: g_dust approx}) because for the IR wavelength dust absorption is typically negligible.
For a dust absorption cross section of $4.5 \times 10^{-23}$ cm$^2$ per hydrogen nucleus (for the 2–3 $\mu$m wavelength range; \citealt{Draine2011}), the gas remains optically thin up to gas column densities $N \approx 2 \times 10^{22}$ cm$^{-2}$.

The factor $\mathcal{N}^{\rm IR} = \mathcal{I}_{\rm tot}^{\rm IR}/\mathcal{I}_{\rm tot}$ is the ratio of IR to FUV photons emitted per H$_2$ photo-excitation. 
This factor arises because IR emission involves cascades through multiple rovibrational states, unlike FUV emission where each excitation produces a single photon. For instance, when the excited H$_2$ state $B^1\Sigma^+_u$ decays to $X^1\Sigma^+_g$, it emits one FUV photon and leaves H$_2$ rovibrationally excited, which may then emit multiple IR photons (e.g., three photons in the path $v=3 \rightarrow 2 \rightarrow 1 \rightarrow 0$). 
Using the Meudon PDR code data \citep{LePetit2006}, we calculated $\mathcal{N}^{\rm IR}$ for various gas temperatures, considering photo-pumping from $X^1\Sigma^+_g$ to excited levels within $B^1\Sigma^+_u$ and $C^1\Pi^u$, assuming a Boltzmann distribution for $X^1\Sigma^+_g$ rovibrational states and a \citet{Draine1978} FUV radiation spectrum. Using Einstein $A$ coefficients, we computed decay probabilities and the resulting FUV and IR line emissions. For $T=100$ K, we found $\mathcal{N}^{\rm IR} = 3.6$, a value that varies little (standard deviation 0.09, min-max variation 0.3) for temperatures between 10 and 1000 K, indicating that $\mathcal{N}^{\rm IR}$ remains relatively constant across a wide range of temperatures relevant to molecular clouds.

While in terms of photon number, the total IR line emission is higher, the energy surface brightness (erg cm$^{-2}$ s$^{-1}$ sr$^{-1}$) is greater in the FUV, due to the higher energies carried by the FUV photons.

\subsection{Additional H$_2$ destruction processes}
\label{sub: discuasion - CRs}
Our model focuses on H$_2$ photodissociation, neglecting other destruction mechanisms such as cosmic ray (CR) ionization and dissociation, X-ray ionization, and collisional dissociation in warm/hot gas. It is thus best suited for standard molecular clouds that are typically cold ($\ll 10^4$ K) and not exposed to abnormally strong X-ray or CR fluxes. In Appendix \ref{app: CRs}, we describe a detailed model explicitly calculating the effect of CR ionization on H$_2$ removal rate compared to FUV photodissociation. 
For a typical CR ionization rate of $\zeta_0=10^{-16}$ s$^{-1}$, CR destruction remains negligible compared to photodissociation for lines of sight with column densities $N \lesssim 2 \times 10^{22}$ cm$^{-2}$ (see Appendix \ref{app: CRs} and Fig.~\ref{fig: Diss rate H2 FUV vs CRs}), consistent with analytic model predictions \citep{Sternberg2024}. X-rays similarly affect H$_2$ destruction, producing secondary electrons that lead to H$_2$ ionization and dissociation analogous to CRs. 

In regions with abnormally high CR or X-ray fluxes, an apparent imbalance between H$_2$ formation and dissociation rates (as measured by FUV emission lines) may be observed, even if H$_2$ is in CSS, because FUV emission lines only reflect the contribution from FUV excitation and photo-dissociation, not the total H$_2$ removal rate (although at sufficiently high fluxes or column densities, CRs may also contribute to FUV line emission; \citealt{Padovani2024}). For such clouds, the additional contribution of CR ionization and dissociation to H$_2$ removal can be constrained by observing various molecular ions (e.g., H$_3^+$, OH$^+$, H$_2$O$^+$, ArH$^+$) in absorption spectroscopy \citep[e.g.,][]{VanderTak2000, Indriolo2012, Neufeld2017b, Bialy2019c}. Alternatively, the CR contribution may be derived from H$_2$ observations by targeting specific IR emission lines (within the 2-3 $\mu$m range) produced by CR-excited H$_2$, with the relative contribution of CR-excited versus UV-excited H$_2$ lines constrained by line ratios \citep{Bialy2020, Padovani2022, gaches2022}. These lines may be observable with the NIRSpec spectrograph on JWST or, for exceptionally bright CR fluxes, by ground-based observatories \citep{Bialy2022a, Bialy2024}.

\subsection{Observations of H$_2$ in the FUV and IR}
\label{sub: diss UV vs IR}

The H$_2$ IR emission lines can be observed with JWST's complementary instruments: NIRSpec and MIRI. NIRSpec \citep{jakobsen2022} offers low resolution ($R=100$) prism spectroscopy over the entire wavelength range $\lambda=0.6-5.3$~$\mu$m, as well as medium-to-high resolution ($R=1000-2700$) gratings covering various wavelength intervals within this range. The MIRI instrument \citep{Bouchet2015} extends the observable H$_2$ IR line emission spectrum to longer wavelengths.

No FUV instrument has had or currently has the ability to resolve the H$_2$ fluorescent lines over the large solid angles spanned by MCs.
FUV H$_2$ emission from individual protoplanetary disks was measured by HST COS \citep{Herczeg2004,Herczeg2006} and rocket-borne experiments \citep{Hoadley2014,Hoadley2016}.
Low spatial and spectral resolution measurements spanning 70\% of the sky with the FIMS/SPEAR mission \citep{Jo2017} showed intense H$_2$ emission from star-forming regions across our Galaxy. However, the spectral resolution of less than 1,000 was too low to separate individual fluorescent lines and determine the excitation conditions, while the spatial resolution of about $5^\prime$ was coarser than the clouds' scales of variation, leading to great uncertainty in characterizing the H$_2$ formation and dissociation rates in even the nearest clouds.

The lack of FUV observations of H$_2$ has inspired several mission concepts. Building on a well-received but not selected initial proposal for a Medium Explorer-class space telescope \citep{Hamden2022}, we have refined the concept to match NASA's Small Explorer opportunity. This revised mission, named \textit{Eos}, is designed to measure FUV lines from nearby star-forming regions with spectral resolution $R>10,000$, sufficient to distinguish individual fluorescent lines. \textit{Eos} will achieve an angular resolution of $<$ 10 arcseconds over a spectrograph slit several degrees long, enabling a detailed resolution of cloud structures while providing coverage adequate to assess global evolutionary states. 
The \textit{Eos} telescope will implement the measurement approach outlined in this paper, covering thousands of square degrees on the sky during a 2-year primary mission, and surveying nearly all nearby molecular clouds. A brief description of some of the mission objectives can be found in \citet{2024Hamden}. \textit{Eos} will, for the first time, determine the extent of MCs out of CSS and address fundamental questions about cloud origins, evolution, and dispersal in their role as stellar nurseries. \textit{Eos} will be proposed in 2025 to the expected NASA Small Explorer announcement of opportunity, with a launch date in the early 2030s.

\section{Conclusions}
\label{sec:con}

In this study, we have investigated the photodissociation and formation processes of H$_2$ in simulated molecular clouds (MCs), with a particular focus on utilizing FUV and IR line emissions to constrain these rates. Our key findings are as follows:
\begin{enumerate}
    \item A significant fraction of MC mass and volume is out of chemical steady state (CSS) due to rapid dynamical evolution, with H$_2$ formation rates either exceeding or lagging behind photodissociation rates (Tables~\ref{tab:table sim properties}-\ref{tab:rates_comparison}).
    \item The total intensity of H$_2$ line emission, whether measured in FUV or IR, can effectively constrain the line-of-sight (LOS) integrated H$_2$ photodissociation rate $\dot{\Sigma}_{D}$ (Eqs.~\ref{eq: D-Itot}, \ref{eq: D-Itot IR}).
    
    \item Measurements of HI and total gas column density provide a means to constrain the integrated H$_2$ formation rate, $\dot{\Sigma}_{F}$ (Eq.~\ref{eq: JF observation 2}).
    
    \item By combining H$_2$ line emission and gas column densities, we can determine the chemical state of gas along the LOS: CSS ($\dot{\Sigma}_{F}/\dot{\Sigma}_{D} \approx 1$), active H$_2$ formation ($\dot{\Sigma}_{F}/\dot{\Sigma}_{D} \gg 1$), or active photodissociation ($\dot{\Sigma}_{F}/\dot{\Sigma}_{D} \ll 1$) (Eq.~\ref{eq: JF_to_JD}).
    
    \item CSS assessment reveals key aspects of MC evolution. MCs far from CSS suggest recent rapid changes - either fresh gas inflow or quick evaporation. These changes happen faster than the time needed for chemical balance (Eq.~\ref{eq: t chem}).

\end{enumerate}

These findings provide valuable insights into the dynamics of H$_2$ in MCs and offer observational strategies to probe the chemical and physical states of these complex systems. Our results underscore the importance of considering non-steady-state conditions in MC studies and highlight the potential of using molecular line emissions as diagnostic tools for cloud evolution.

\vspace{0.2 cm}
\noindent
SB acknowledges financial support from the physics department at the Technion, and from the Center for Theory and Computational (CTC) at the University of Maryland College Park.
DS and SW thank the Deutsche Forschungsgemeinschaft (DFG) for funding through the SFB~956 ``The conditions and impact of star formation'' (sub-projects C5 and C6). Furthermore, DS and SW receive funding from the programme ``Profilbildung 2020'', an initiative of the Ministry of Culture and Science of the State of North Rhine-Westphalia.
TJH is funded by a Royal Society Dorothy Hodgkin Fellowship and UKRI ERC guarantee funding (EP/Y024710/1). 
B.B. acknowledges support from NSF grant AST-2009679 and NASA grant No. 80NSSC20K0500.
B.B. is grateful for generous support by the David and Lucile Packard Foundation and Alfred P. Sloan Foundation. 
The work was carried out in part at the Jet Propulsion Laboratory, California Institute of Technology, under a contract with the National Aeronautics and Space Administration.
The calculations were carried out using the Numpy and Scipy libraries \citep{numpy, scipy}
The figures were produced using the matplotlib library \citep{matplotlib}. The interactive figure was produced using Plotly and Github.

\appendix
\section{The $\dot{\Sigma}_{D}-\mathcal{I}_{\rm tot}$ relation}
\label{app: Sigma-Itot}

We derive the relationship between the integrated H$_2$ photodissociation rate, $\dot{\Sigma}_{D}$, and the total H$_2$ line emission intensity, $\mathcal{I}_{\rm tot}$.
Following \citetalias{Sternberg1989a}, the total photon intensity for a uniform density 1D slab is given by
\begin{align}
\label{app eq: Itot 1}
    \mathcal{I}_{\rm tot} & = \frac{1}{4 \pi} (1-p_{\rm diss}) \int_0^L n({\rm H_2}) \chi P_0 \mathrm{e}^{-2\tau} f_{\rm H_2, shield} \mathrm{d}s \\
    &= \frac{1}{4 \pi} (1-p_{\rm diss}) \int_0^L n({\rm H_2}) P \mathrm{e}^{-\tau} \mathrm{d}s  
    \ .
    \label{app eq: Itot 2}
\end{align}
To derive this relation, we start with Eq.~(A1) in \citetalias{Sternberg1989a}, which describes the intensity of a single H$_2$ emission line. We sum over the branching ratios (denoted as $b$ in \citetalias{Sternberg1989a}) for transitions to bound rovibrational states, introducing the factor $(1-p_{\rm diss})$ in our Eqs.~(\ref{app eq: Itot 1}-\ref{app eq: Itot 2}). In these equations, $P_0$ is the un-attenuated photo-excitation rate for a unit Draine \citep{Draine1978} radiation field, while $P$ is the local, attenuated H$_2$ photo-excitation rate at cloud depth $s$. For slab geometry, $P = \chi P_0 f_{\rm H_2, shield} \mathrm{e}^{-\tau}$, where $f_{\rm H_2, shield}$ is the H$_2$ self-shielding function \citep[e.g.,][]{Draine1996}, and $\tau=\int_0^s \sigma n \mathrm{d}s'$ is the dust opacity from point $s$ to the cloud's edge, with $\sigma=1.9\times 10^{-21}$ cm$^2$ being the dust absorption cross-section per hydrogen nucleus averaged over the LW band.

Our notation differs from \citetalias{Sternberg1989a}: our $P_0$ and $D_0$ correspond to their $P$ and $D$, and they denote H$_2$ self-shielding as $f(N_2)$. In Eqs.~(\ref{app eq: Itot 1}-\ref{app eq: Itot 2}), we integrate emission through the cloud (from $s=0$ to $s=L$) along the LOS. The factor of two in the exponent of Eq.~(\ref{app eq: Itot 1}) accounts for dust absorption in both directions: 

first, as LW radiation propagates from the cloud edge inward, exciting H$_2$, and second, as the resulting emission lines travel outward. Unlike FUV continuum radiation, which experiences both dust absorption and H$_2$ self-shielding, H$_2$ emission lines are primarily attenuated by dust alone. 
This is because most H$_2$ in the cloud occupies low ro-vibrational states and thus doesn't re-absorb these emission lines which mostly correspond to high rovibrational lower states. However, this approach is an approximation, as some self-absorption can occur, and may alter the H$_2$ emission spectrum at shorter wavelengths (Le Bourlot, private communication).

Examining Eq.~(\ref{app eq: Itot 2}), we notice its similarity to $\dot{\Sigma}_D^{\rm (true)}$ (Eq.~\ref{eq: JF JD}), differing only by a constant multiplication term and an exponential term in the integral. To express $\mathcal{I}_{\rm tot}$ in terms of $\dot{\Sigma}_D^{\rm (true)}$, we first define an effective attenuation factor:
\begin{equation}
\label{app eq: g_dust}
   \beta_{\rm dust} \equiv \frac{\int n({\rm H_2}) D  \mathrm{e}^{-\tau} \mathrm{d}s}{\int n({\rm H_2}) D \mathrm{d}s} \ .
\end{equation}
Using this definition, we can combine Eq.~(\ref{app eq: Itot 2}) with Eq.~(\ref{eq: JF JD}) to derive a relation between the H$_2$ line intensity and the integrated H$_2$ photodissociation rate:
\begin{equation}
\label{app eq: D-Itot v0}
    \mathcal{I}_{\rm tot} = \frac{1-p_{\rm diss}}{4 \pi p_{\rm diss} \bar{m}} \beta_{\rm dust} \ \dot{\Sigma}_D^{\rm (true)} \ .
\end{equation}
In this derivation, we utilize the relationship  $P=D/p_{\rm diss}$ (Eq.~\ref{eq: p_diss}).

The $\beta_{\rm dust}$ factor accounts for the reduction in line emission due to dust absorption as photons propagate from the cloud interior to the observer. However, direct measurement of $\beta_{\rm dust}$ is challenging as it depends on the 3D density structure along the LOS and the radiation geometry. While 3D dust maps provide information on density structure (e.g., \citealt{Leike2020a}), they typically cannot resolve the critical HI-to-H$_2$ transition length, which usually occurs over scales $\lesssim 1$~pc \citep{Bialy2017}.
In the absence of 3D information, we approximate $\beta_{\rm dust}$ using the simplest possible geometry: a 1D uniform slab where H$_2$ line emission and dust absorption occur, with total dust optical depth $\tau_{\rm tot} = \int_0^L \sigma n ds =\sigma N$, where $N$ is the integrated gas column density along the entire LOS. Assuming equal line emission per unit dust optical depth $d\tau$, we can express $\beta_{\rm dust}$ as:
 \begin{equation}
 \label{app eq: g_dust approx}
     \beta_{\rm dust}^{\rm (obs)} = \int_0^{\tau_{\rm tot}} \left(\frac{\mathrm{d}\tau}{\tau_{\rm tot}} \right) \mathrm{e}^{-\tau}  = \frac{1-\mathrm{e}^{-\tau_{\rm tot}}}{\tau_{\rm tot}} = \frac{1-\mathrm{e}^{-1.9 N_{21}}}{1.9 N_{21}}  \ .
 \end{equation}
This expression is the exact solution given by the equation of transfer for attenuation in a medium where emitters are fully mixed with absorbers, and is analogous to the escape probability formalism. In the last equality, we use $\sigma=1.9 \times 10^{-21}$ cm$^{-2}$ and define $N_{21} \equiv N/(10^{21} \ {\rm cm^{-2}})$, yielding $\tau_{\rm tot}=1.9 N_{21}$. 
In the limit $\tau_{\rm tot} \ll 1$, $\beta_{\rm dust, obs} \rightarrow 1$, approaching the optically thin limit. Conversely, when $\tau_{\rm tot} \gg 1$, $\beta_{\rm dust, obs} \rightarrow 1/\tau_{\rm tot}$. This is because in the optically thick limit, the LOS contains $\sim \tau_{\rm tot}$ transition layers (each with opacity $\sim 1$), but we receive signals only from emission lines to a depth of $\tau_{\rm tot} \sim 1$, resulting in relative emission $\propto 1/\tau_{\rm tot}$.

Substituting Eq.~(\ref{app eq: g_dust approx}) for $\beta_{\rm dust}$ in Eq.~(\ref{app eq: D-Itot v0}), we obtain:
\begin{align}
\label{app eq: D-Itot}
    \dot{\Sigma}_{D}^{\rm (obs)} &=  \frac{4 \pi p_{\rm diss} \bar{m}}{1-p_{\rm diss}} \ \mathcal{I}_{\rm tot}  \   \left(\frac{\tau_{\rm tot}}{1-\mathrm{e}^{-\tau_{\rm tot}}} \right)  \\ \nonumber 
    &= 0.3 \ \mathcal{I}_5 \left( \frac{N_{21}}{1-\mathrm{e}^{-1.9 N_{21}}} \right) \ {\rm M_{\odot} \ pc^{-2} \ Myr^{-1}} \ ,
\end{align}
where in the numerical evaluation we use $\tau_{\rm tot}=1.9 N_{21}$ and define $\mathcal{I}_5 \equiv \mathcal{I}_{\rm tot}/({10^{5} \ {\rm photons \ cm^{-2} \ s^{-1} \ sr^{-1}}})$. The superscript ``(obs)" in Eqs.~(\ref{app eq: g_dust approx}) and (\ref{app eq: D-Itot}) emphasizes that these expressions provide approximations to the true attenuation factor and H$_2$ photodissociation rates, as they rely on observable (integrated) quantities.

\begin{figure*}
    \centering
    \includegraphics[width=1.0\textwidth]{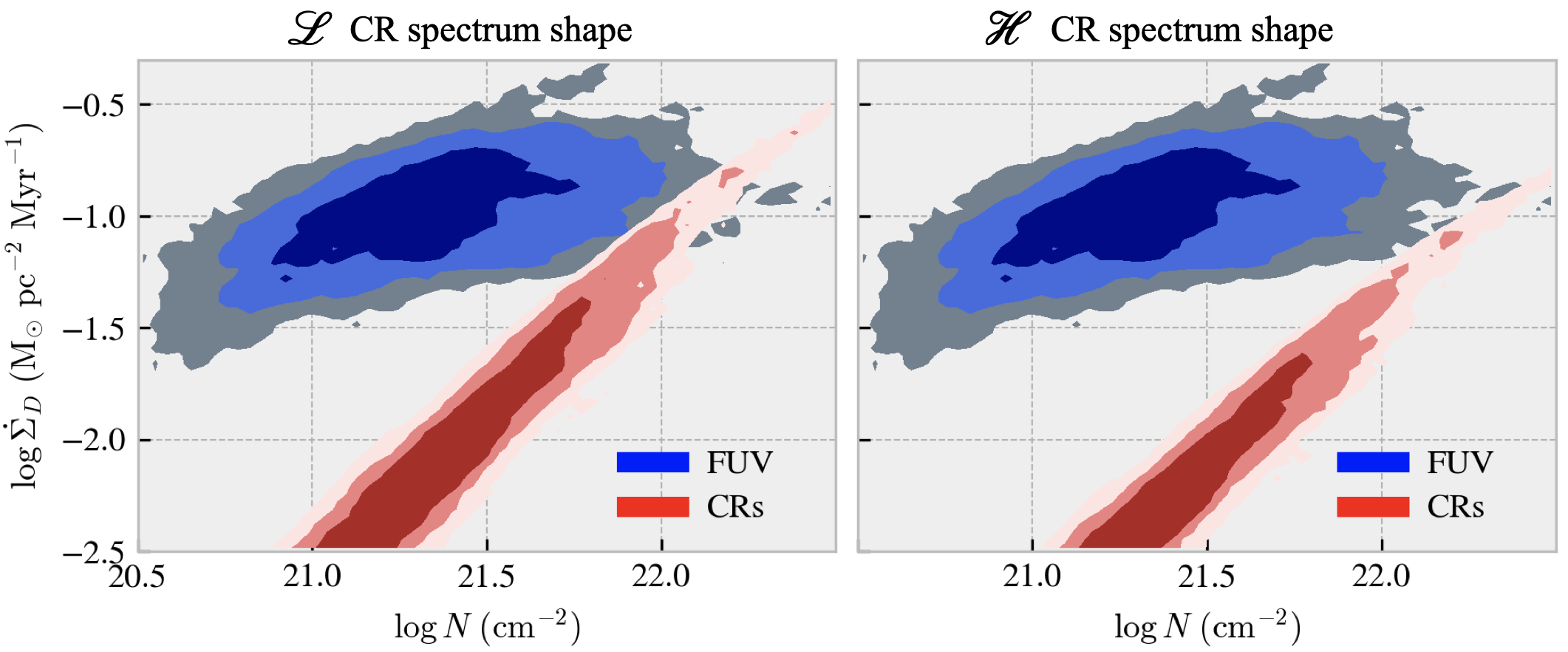}
    \caption{Contributions of FUV radiation and CRs to H$_2$ removal as a function of LOS column density. The left panel shows results for the $\mathcal{L}$ CR spectral shape model, while the right panel represents the $\mathcal{H}$ model \citep[these models are discussed in][]{Padovani2018}. Both assume an unattenuated CR ionization rate of $\zeta_0=10^{-16}$ s$^{-1}$ and  $N_0=10^{19}$ cm$^{-2}$ (see Eq.~\ref{eq: D zeta}-\ref{eq: zeta}). Shaded regions represent contours of the 2D PDF of $\log \dot{\Sigma}_D$ versus $\log N$, with light to dark shades corresponding to PDF/PDF$_{\rm max}$ levels that fall within the ranges (0.03, 0.1), (0.1, 0.3), and (0.3, 1), respectively. The figure demonstrates that for column densities $N \lesssim 2 \times 10^{22}$ cm$^{-2}$, FUV photodissociation dominates over CR-induced H$_2$ removal, consistent with \citet{Sternberg2024}.}
    \label{fig: Diss rate H2 FUV vs CRs} 
\end{figure*}
\section{H$_2$ destruction by cosmic rays}
\label{app: CRs}
To assess the contribution of cosmic rays (CRs) to H$_2$ removal, we calculated the individual contributions of FUV and CRs to H$_2$ dissociation on a cell-by-cell basis in our simulation. In each cell, the H$_2$ removal rate (s$^{-1}$) by CRs is:
\begin{equation}
\label{eq: D zeta}
    D_{\rm CR} = \varphi \zeta
\end{equation}
where $\zeta$ (s$^{-1}$) is the total H$_2$ ionization rate by primary CRs and secondary electrons, and $\varphi$ is the number of H$_2$ molecules destroyed per ionization event. This includes H$_2$ ionization by CRs, H$_2$ dissociation via chemical reactions with CR-produced ions, and direct H$_2$ dissociation by CRs. Following \citet{Sternberg2024}, we adopt $\varphi=2$.
The CR ionization rate is given by:
\begin{equation}
\label{eq: zeta}
    \zeta = 
    \begin{cases}
    \zeta_0 & \text{if } N_{\rm eff}<N_0 \\ 
    \zeta_0 \left( \frac{N_{\rm eff}}{N_0} \right)^{-\gamma} & \text{if } N_{\rm eff} \geq N_0
    \end{cases}
\end{equation}
This form accounts for CR attenuation as they propagate through the cloud, where $N_{\rm eff}$ represents an effective mean column density for CRs propagating in different directions, and $\zeta_0$, $N_0$ and $\gamma$ are constants. We approximate $N_{\rm eff}$ using the effective visual extinction $A_{V, {\rm eff}}$ as calculated by the SILCC simulation, with $N_{\rm eff} = 1.8 \times 10^{21} A_{V, {\rm eff}}$ cm$^{-2}$ \citep[see][for a discussion of this approach]{gaches2022}.
We consider the two CR energy spectral shape models $\mathcal{H}$ and $\mathcal{L}$ discussed by \citet{Padovani2018}, for which $\gamma=0.39$ and 0.28, respectively and adopt the standard values $N_0=10^{19}$ cm$^{-2}$ and $\zeta_0=10^{-16}$ s$^{-1}$ \citep{Sternberg2024}. 
Using these equations, we calculate the H$_2$ removal rate by CRs on a cell-by-cell basis. We then integrate these rates along the LOS (Eq.~\ref{eq: JF JD}) to obtain the CR contribution to the H$_2$ integrated removal rate, $\dot{\Sigma}_D$.

In Fig.~\ref{fig: Diss rate H2 FUV vs CRs}, we present the FUV and CR contributions to $\dot{\Sigma}_D$ as a function of the column density along the LOS, $N$. The left and right panels correspond to the $\mathcal{L}$ and $\mathcal{H}$ CR spectral shape models, respectively. 
The three shaded regions in each panel represent contours of the 2D PDF of $\log \dot{\Sigma}_D$ versus $\log N$. These contours correspond to PDF/PDF$_{\rm max}$ levels in the ranges (0.03, 0.1), (0.1, 0.3), and (0.3, 1), depicted by light to dark shades, respectively.
Our analysis shows that for LOS with column densities $N \lesssim 2 \times 10^{22}$ cm$^{-2}$, the CR contribution to H$_2$ removal is negligible compared to FUV photodissociation. This finding is consistent with the analytic model predictions of \citet{Sternberg2024}.
It is important to note that our chosen value of $\zeta_0=10^{-16}$ s$^{-1}$ represents a standard CR ionization rate in the ambient ISM. However, in the vicinity of CR sources, such as supernova remnants, the CR flux may be significantly higher, resulting in an increased value of $\zeta_0$. In such cases, the CR distribution shown in our figure would be shifted upward in proportion to the increase in $\zeta_0$. Consequently, the transition point where CR-induced H$_2$ removal becomes comparable to FUV photodissociation would occur at lower column densities.

\section{Statistical Analysis of Multiple Snapshots}
\label{app: snapshots stats}

To validate the robustness of our methodology across different cloud conditions, we have analyzed multiple snapshots of our simulation. Specifically, we examined snapshots at $t = 2, 4$, and $5$ Myr, in addition to our original snapshot at $t = 3$ Myr. For each time, we analyzed maps along three orthogonal LOS, $x$, $y$, and $z$, resulting in a total of 12 realizations.

For each realization, we calculated the integrated dissociation and formation mass rates ($\dot{M}_D$ and $\dot{M}_F$) by integrating their corresponding surface density rates ($\dot{\Sigma}_{D}$ and $\dot{\Sigma}_{F}$) over the area. 
Specifically, integrating the observationally-derived surface densities $\dot{\Sigma}_{D}^{\rm (obs)}$ and $\dot{\Sigma}_{F}^{\rm (obs)}$ using Eqs.~(\ref{eq: D-Itot}) and (\ref{eq: JF observation 2}) yields the observationally-derived mass rates $\dot{M}_D^{\rm (obs)}$ and $\dot{M}_F^{\rm (obs)}$, while integrating the true surface densities $\dot{\Sigma}_{D}^{\rm (true)}$ and $\dot{\Sigma}_{F}^{\rm (true)}$ using Eq.~(\ref{eq: JF JD}) yields the true mass rates $\dot{M}_D^{\rm (true)}$ and $\dot{M}_F^{\rm (true)}$.
The true rates can also be calculated (identically) by volumetrically integrating the H$_2$ formation and dissociation volume rates $j_F$ and $j_D$ (Eq.~\ref{eq: jf jd}) over the simulation volume.
Since the true mass rates represent cloud-integrated quantities over the full volume, they are independent of orientation ($x$, $y$, $z$), while the observationally-derived rates depend on the viewing angle as they are estimated from 2D projected maps. 
We report these values in Table~\ref{tab:rates_comparison}, and compare them via the absolute relative difference:
\begin{align}
\label{eq: delta}
\Delta_D &= \frac{|\dot{M}_D^{\rm (obs)}-\dot{M}_D^{\rm (true)}|}{\dot{M}_D^{\rm (true)}} \nonumber \\
\Delta_F &= \frac{|\dot{M}_F^{\rm (obs)}-\dot{M}_F^{\rm (true)}|}{\dot{M}_F^{\rm (true)}} \ .
\end{align}
Across all 12 realizations, we find mean relative differences of $\bar{\Delta}_D = 0.31$ and $\bar{\Delta}_F = 0.27$, with median values of 0.29 and 0.18 for dissociation and formation, respectively (see Table~\ref{tab:rates_comparison} for more details).

\begin{table*}
\caption{H$_2$ Formation and Dissociation Rates for Multiple Snapshots}
\label{tab:rates_comparison}
\begin{center}
\begin{tabular}{c | c c c | c c c}
\toprule
Time & $\dot{M}_F^{\rm (true)}$ & $\dot{M}_F^{\rm (obs)}$ & $\Delta_F$ & $\dot{M}_D^{\rm (true)}$ & $\dot{M}_D^{\rm (obs)}$ & $\Delta_D$ \\
(Myr) & \multicolumn{2}{c}{($10^3~\mathrm{M_\odot~Myr^{-1}}$)} & & \multicolumn{2}{c}{($10^3~\mathrm{M_\odot~Myr^{-1}}$)} & \\
\midrule
2 & 7.46 & 13.3 & 0.78 & 0.965 & 1.23 & 0.28 \\
    &      & 8.90  & 0.19 &       & 1.42 & 0.48 \\
    &      & 7.80  & 0.05 &       & 1.30 & 0.34 \\
\midrule
{\bf 3} & 8.61 & 9.30  & 0.08 & 1.14 & 1.41 & 0.24 \\
{\bf (fiducial)}    &      & 10.8 & 0.26 &      & 1.47 & 0.29 \\
    &      & 7.26  & 0.16 &      & 1.07 & 0.06 \\
\midrule
4 & 10.1 & 15.5 & 0.54 & 1.03 & 1.33 & 0.30 \\
    &      & 9.15  & 0.09 &      & 1.47 & 0.43 \\
    &      & 7.07  & 0.30 &      & 1.31 & 0.28 \\
\midrule
5 & 8.56 & 14.2 & 0.66 & 1.04 & 1.38 & 0.32 \\
    &      & 8.47  & 0.01 &      & 1.49 & 0.43 \\
    &      & 7.52  & 0.12 &      & 1.32 & 0.27 \\
\bottomrule
\end{tabular}
\end{center}
\textbf{Notes.} For each snapshot time, we show the true rates from the simulation ($\dot{M}^{\rm (true)}$) and observationally-derived rates ($\dot{M}^{\rm (obs)}$) for H$_2$ formation and dissociation. Each snapshot contains three rows corresponding to orthogonal viewing angles (x, y, and z LOS orientations). The relative differences $\Delta$ are calculated using Eq.~(\ref{eq: delta}). Our fiducial snapshot ($t=3$ Myr) that is analyzed throughout the paper is highlighted in boldface. Statistics shown at the bottom are computed across all measurements (12 realizations: 4 snapshots × 3 LOS orientations).
\end{table*}

\end{document}